\documentclass[aps,pre,twocolumn,groupedaddress]{revtex4-1}
\usepackage{graphicx}
\usepackage{amssymb}
\usepackage{amsmath}
\usepackage{bm}
\usepackage[caption=false]{subfig}
\newcommand{\tdp}{t^{\prime}}
\begin{document}
\title{Emergence of new topological gapless phases in the modified square-lattice Kitaev model}

\author{Jihyeon \surname{Park}}
\affiliation{Department of Physics, Ewha Womans University, Seoul 03760, Korea}
\author{Gun Sang \surname{Jeon}}
\email[gsjeon@ewha.ac.kr]{}
\affiliation{Department of Physics, Ewha Womans University, Seoul 03760, Korea}


\begin{abstract}
We investigate emergent topological gapless phases in the square-lattice
		  Kitaev model with additional hopping terms. 
		  In the presence of nearest-neighbor hopping only, 
		  the model is known to exhibit gapless phases with two
		  topological gapless points. 
		  When the strength of the newly added next-nearest-neighbor hopping is smaller
		  than a certain value, qualitatively the same phase diagram persists. We find that
		  further increase of the extra hopping results in a new topological phase
		  with four gapless points. 
		  We construct a phase diagram to clarify the regions of emergent
		  topological gapless phases as well as topologically trivial ones
		  in the space of the chemical potential and the next-nearest-neighbor hopping strength. 
		  We examine the evolution of
		  the gapless phases in the energy dispersions of the bulk as 
		  the chemical potential varies. 
		  The topological properties of the
		  gapless phases are characterized by the winding numbers of the present gapless points. 
		  We also consider the ribbon geometry to examine the corresponding
		  topological edge states. 
		  It is revealed that Majorana-fermion edge modes exist as flat bands in topological gapless
		  phases.
		  We also perform the analytical calculation as to Majorana-fermion
		  zero-energy modes and discuss its implications on the numerical results.
\end{abstract}

\flushbottom
\maketitle
\thispagestyle{empty}

\section{Introduction}

Topological properties of condensed matter have emerged as a new
paradigm for classifying materials~\cite{Hasan2010, Wen1995}. 
They have attracted intense interests in view of their robustness against any continuous
deformation. 
Topological insulators and topological superconductors are prototypes
for 
interesting topological materials which have topologically nontrivial gapped
phases. Topological insulators exhibit an insulating gap in the bulk and host
gapless surface states preserved by time-reversal symmetry~\cite{Hasan2010}.
Topological superconductors have a pairing gap in the bulk and remarkably
manifest Majorana particles on the surface attributed to particle-hole symmetry
in the system~\cite{Qi2011, DasSarma2023}.  

Majorana fermion is a particle which is its own antiparticle, suggested as real
solutions of the Dirac equation by Ettore Majorana~\cite{Alicea2012, Beenakker2013,DasSarma2023}. 
In high energy physics neutrinos may be a candidate for Majorana
fermions and experiments to demonstrate it have been still ongoing. 
Majorana particles are also of interest in condensed matter physics and they are
proposed to be
created as quasiparticles which are a linear
combination of electrons and holes with equal weights~\cite{Alicea2012, Beenakker2013, Stanescu2013,
Elliott2015, Sato2016, DasSarma2023}. 
Researches on the realization of Majorana fermions 
have been performed intensively since Majorana fermions are promising
candidates for topological qubits in fault-tolerant quantum computing. 
The
experimental realization of Majorana modes in various platforms of
superconductor has been under intensive researches but
the obvious manifestation of Majorana edge modes is still in controversy~\cite{Law2009,Mourik2012,He2017,Zhang2018b,Nayak2008,DasSarma2023}.  

Theoretically
it was shown that one-dimensional superconductor can
host unpaired Majorana zero modes at the ends~\cite{Kitaev2001}. 
The existence of
the Majorana zero modes depends on $\mathbb{Z}_{2}$-type topological invariant
of the one-dimensional superconductor.
Two-dimensional
topological superconductors in $\mathbb{Z}$-type integer classification host
Majorana chiral edge modes and the integer topological invariant indicates the
number of the edge modes. 
The simplest of these models is the
spinless two-dimensional superconductors with $p_{x}+ip_{y}$ symmetry. 
This
model hosts Majorana zero modes at the core of the vortex as well as Majorana
chiral modes on the boundaries~\cite{Read2000}. 

The search for Majorana zero modes has been conducted in topologically trivial
superconductors as well.
Majorana zero modes have been
proposed to be realized in a two-dimensional superconducting Dirac semimetal
with trivial bulk topology~\cite{Chan2017}.   
It was also pointed out that a Hopf invariant can protect Majorana zero modes
in superconductors without chiral edge modes~\cite{Yan2017}.
It has also been shown that
two-dimensional topological insulators can exhibit Majorana zero modes at each
corner
when they are in the proximity with $d$-wave~\cite{Yan2018} or
$s_\pm$-wave~\cite{Wang2018a} superconductors.
Topological gapless phases, which occur in topologically trivial
superconductors described by the square-lattice Kitaev model, 
were also reported to display Majorana fermions in the
form of a flat band~\cite{Wang2017a,Zhang2019}.
At the edges such flat bands connect two gapless points of opposite chirality in the
bulk~\cite{Ryu2002,Delplace2011,Chiu2014,Zhang2020} which are
preserved by the topological invariant of the model~\cite{Matsuura2013}.

In this work, we extend the square-lattice Kitaev model to explore the
possibility of newly emergent gapless phases. 
For this purpose, we employ the modified square-lattice Kitaev model
by adding next-nearest-neighbor hopping terms to the original model.
In the earlier study~\cite{Wang2017a}, the model with extra diagonal hopping 
was considered, leading to the conclusion that it does not give rise to
any qualitatively different topological feature.
We here find that
topological gapless phases of a new type emerge 
from the introduction of extra next-nearest-neighbor hopping when it is imposed
in the other diagonal direction.
We perform the detailed analysis of the new topological phases in view of the phase
boundary, the bulk energy dispersions, and Majorana zero-energy flat bands in the ribbon
geometry. 
The rich physics of Majorana fermions in the modified square-lattice Kitaev
model is also revealed in an analytical way.

\section{Modified square-lattice Kitaev model}

\subsection{Model}

Square-lattice Kitaev model represents two-dimensional spinless superconductors
with $p$-wave symmetry\cite{Li2006,Wang2017a,Zhang2019}, which is a
two-dimensional generalization of the one-dimensional spinless fermionic
Hamiltonian\cite{Lieb1961,Kitaev2001}. We consider the modified square-lattice
Kitaev model which includes additional second nearest neighbor hopping terms.
The Hamiltonian of the modified square-lattice Kitaev model is given by  

\begin{align} 
	\begin{split}
	H =& -t \sum_{\bm{r},\bm{a}} \left( c^{\dagger}_{\bm{r}} c_{\bm{r}+\bm{a}} + c^{\dagger}_{\bm{r}+\bm{a}} c_{\bm{r}} \right)  
	-\Delta\sum_{\bm{r},\bm{a}} \left( c_{\bm{r}} c_{\bm{r}+\bm{a}} + c^{\dagger}_{\bm{r}+\bm{a}} c^{\dagger}_{\bm{r}}\right) \\
	&	-\tdp \sum_{\bm{r}} \left( c^{\dagger}_{\bm{r}} c_{\bm{r}+\bm{b}} +c^{\dagger}_{\bm{r}+\bm{b}} c_{\bm{r}} \right)
	+\mu \sum_{\bm{r}} \left(2c^{\dagger}_{\bm{r}} c_{\bm{r}}-1\right)\label{H},
	\end{split}
\end{align}
where $c^{\dagger}_{\bm{r}}$($c_{\bm{r}}$) is the creation(annihilation)
operators of an electron at site $\bm{r}$ on the square lattice of $N$
lattice points.
The first summation represents the hopping of strength $t$ between nearest neighbor
sites.
The second summation denotes Cooper pairing with $p$-wave symmetry, where $\Delta$
is an isotropic order parameter assumed to be real. 
The additional hopping terms of hopping amplitude $\tdp$ are introduced in the
third summation, where $\bm{b}\equiv a\bm{\hat{i}}-a\bm{\hat{j}}$ with $a$ being
the lattice constant and $\bm{\hat{i}}(\bm{\hat{j}})$ being a unit vector in
$x\left( y\right)$ direction. The last summation arises from chemical potential
$\mu$. 
Throughout the paper 
We 
write all the energy quantities and the length
scales in units of $t$ and $a$, respectively.  
 
In order to examine the bulk phase we assume the periodic boundary conditions in both $x$ and $y$ directions. Taking the Fourier transformation on the fermion annihilation operators as
%
\begin{align}
		  c_{\bm{r}}=\frac{1}{\sqrt{N}}\sum_{\bm{k}}c_{\bm{k}}e^{i\bm{k}\cdot\bm{r}},
\end{align}
we obtain the Hamiltonian of the form
\begin{align} \nonumber
	H =
	\sum_{\bm{k}}&\bigg[ 2 \big\{ \mu-t\left( \cos{k_{x}}+\cos{k_{y}}\right)-2\tdp\cos(k_{x}{-}k_{y}) 
	\big\} c^{\dagger}_{\bm{k}}c_{\bm{k}}
	\\
	&-i\Delta\left(\sin{k_{x}}+\sin{k_{y}}\right)\left(c^{\dagger}_{\bm{k}}c^{\dagger}_{-\bm{k}}+c_{\bm{k}}c_{-\bm{k}}\right) -\mu\bigg].
\end{align}
It can be expressed in terms of $\Psi_{\bm{k}}\equiv\small( \begin{array}{cc}
	c_{\bm{k}} &  c^{\dagger}_{-\bm{k}}	\end{array}\small)^{\dagger}$
%
\begin{align}
	\nonumber
	H =&
		  \sum_{\bm{k}}\Psi_{\bm{k}}^{\dagger}
		  \bm{h}_{\bm{k}}
		  \Psi_{\bm{k}}
\end{align}
with
\begin{widetext}
	\begin{align}
			  \bm{h}_{\bm{k}} \equiv&
		\begin{pmatrix}
			-\mu+t\left( \cos{k_{x}}+\cos{k_{y}} \right)+\tdp\cos{\left(k_{x}-k_{y}\right)} &  -i\Delta\left(\sin{k_{x}}+\sin{k_{y}}\right)
			\\ i\Delta\left(\sin{k_{x}}+\sin{k_{y}}\right) & \mu-t \left( \cos{k_{x}}+\cos{k_{y}}\right)-\tdp\cos{\left(k_{x}-k_{y}\right)}
		\end{pmatrix}.
	\end{align}
		  The eigenvalues of $\bm{h}_{\bm{k}}$ provide quasiparticle eigenenergies $\pm\epsilon_{\bm{k}}$ with
%
	\begin{align}
		\epsilon_{\bm{k}}\equiv\sqrt{\{-\mu+t \left( \cos{k_{x}}+\cos{k_{y}} \right)+\tdp\cos{\left(k_{x}-k_{y}\right)}\}^2+\Delta^2\left(\sin{k_{x}}+\sin{k_{y}}\right)^2}\label{ev}.
	\end{align}
\end{widetext}
	
\subsection{Construction of phase diagram}

\begin{figure}
	\includegraphics[width=8.0cm,angle=0]{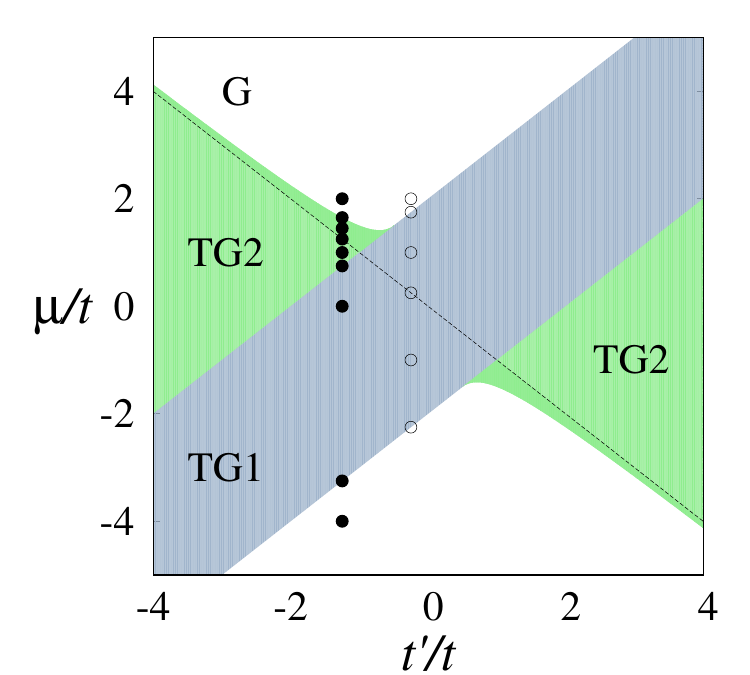}
	\caption{ Phase diagram of the modified square-lattice Kitaev model
		  characterized by the number of gapless points on the 
		  $\mu$-$\tdp$ plane. `G', `TG1', and `TG2' regions indicate gapful phase,
		  gapless phase with a pair of gapless points, and gapless phase with two
		  pairs of gapless points, respectively. 
		  The dotted line on $\mu=-\tdp$ is the
		  boundary where the topological property of gapless points changes. 
		  The empty circles at $\tdp=-0.25t$ and the filled circles at $\tdp=-1.25t$
		  are the representative systems whose properties are presented in
		  subsequent sections.
	}\label{phase_diagram}	
\end{figure}

The position of gapless points are determined by the condition $\epsilon_{\bm{k}}=0$, 
\begin{align}
	&\Delta\left(\sin{k_{x}}+\sin{k_{y}}\right)=0, \label{zero_1}
	\\&\mu-t\left(\cos{k_{x}}+\cos{k_{y}}\right)-\tdp\cos(k_{x}-k_{y})=0. \label{zero_2}
\end{align}
The former equation yields two kinds of locations
\begin{align}
	k_{y}=-k_{x}, \label{kxeq-ky}
	\end{align}
or
	\begin{align}
			  k_y = k_x + \pi, 
			  \label{ky_two_cases}
\end{align}
where $k_x$ and $k_y$ are defined modulo $2\pi$ in the interval $[-\pi,\pi]$.
The insertion of Eq. (\ref{kxeq-ky}) into Eq. (\ref{zero_2}) leads to a quadratic equation of $\cos{k_{x}}$
\begin{align}
	-2\tdp\cos^{2}{k_{x}}-2t\cos{k_{x}}+\mu+\tdp=0, \label{cos_quad}
\end{align}
yielding two formal solutions for $\cos{k_{x}}$
\begin{align}
	\chi_{\pm}\equiv\frac{-t\pm\sqrt{t^{2}+2\tdp\left(\mu+\tdp\right)}}{2\tdp}. \label{sol_cos}
\end{align}
A pair of gapless points
\begin{align}
	k_{x}=-k_{y}=\pm \arccos{\chi_{+}}
\end{align}
are guaranteed to exist in the condition $|\chi_{+}|<1$, which corresponds to the region $ -f(-t') < \mu < f(t')$ with

\begin{align}
		  \displaystyle
  f(x) \equiv 
		  \left\{
					 \begin{array}{cc}
								\displaystyle - x - \frac{t^2}{2x} 
								& \displaystyle \hbox{for } x < - \frac12 t ,
								\\[3mm]
						x + 2t &\displaystyle  \hbox{for } x > - \frac12 t .
			 \end{array}
		 \right.
\end{align}
We can find another pair of gapless points
\begin{align}
	k_{x}=-k_{y}=\pm \arccos{\chi_{-}}
\end{align}
if the system satisfies the condition $|\chi_{-}|<1$; 
the resulting region is
\begin{align}
	\begin{cases}
		\displaystyle t'+2t<\mu<-\frac{t^{2}}{2t'}-t'\quad \displaystyle 
			  &\textrm{for} \quad\displaystyle \tdp<-\frac{1}{2}t,\\
		\displaystyle -\frac{t^{2}}{2t'}-t'<\mu<t'-2t\quad 
			  &\textrm{for} \quad\displaystyle \tdp>\frac{1}{2}t.\label{eq***}
	\end{cases}
\end{align}
Equation~(\ref{ky_two_cases}) combined with Eq.~(\ref{zero_2}) is satisfied on a
line $\mu=-\tdp$, where a nodal line $k_{y}=k_{x} + \pi$ of Eq.~(\ref{ky_two_cases}) exists on the momentum plane. 

Summarizing the above results, we can construct the phase diagram shown in Fig.~\ref{phase_diagram}. 
In the region $\mu > f(\tdp)$ or $\mu < -f(-\tdp)$
the system lies in a normal gapped phase denoted by G in Fig. \ref{phase_diagram}.
In the region $\tdp-2t<\mu<\tdp+2t$
we observe a topological gapless phase with a pair of gapless points (TG1).
In the intermediate region defined in Eq.~(\ref{eq***}) another topological gapless phase with two pairs of gapless points (TG2) emerges.

It is remarkable that the second gapless phase TG2 does not show up in the {\em original} square-lattice Kitaev model even when the second-nearest-neighbor hopping is introduced in the diagonal direction other than that considered in this work.
The system exhibits straight nodal lines on the line $\mu=-\tdp$, which is
inside TG2 for $|\tdp|>t$ and inside TG1 for $|\tdp|<t$. In subsequent sections,
we will give a description to how the phase evolves across the line. 

In the remaining part, we will give a description mostly for $\tdp<0$.
We can always apply the same analysis to the system with $\tdp>0$ by replacing $\mu$ and $\tdp$ by $-\mu$ and
$-\tdp$ in the expressions for $\tdp<0$.

\subsection{Energy dispersions of quasiparticles}

\begin{figure*}
		
	\subfloat[$\mu=2t$]{
			  \includegraphics[width=5.5cm]{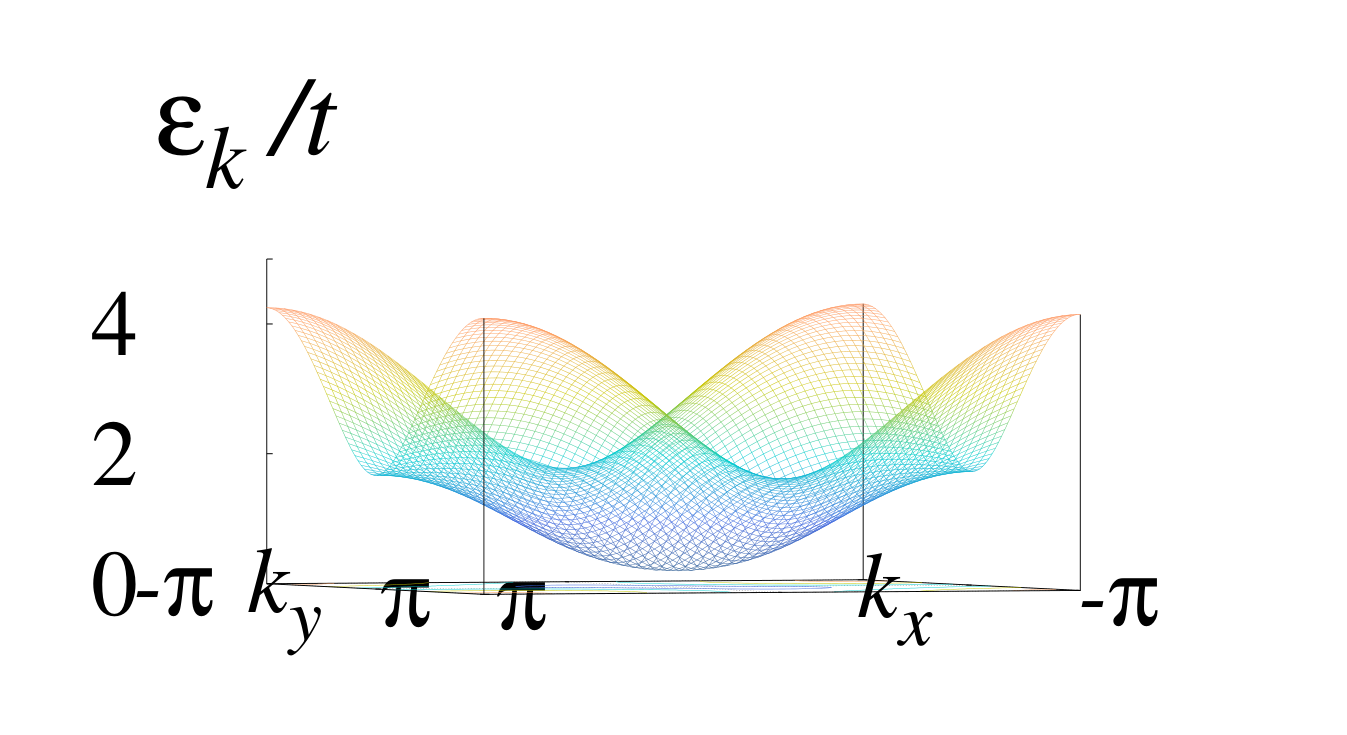}
	}
	\subfloat[$\mu=1.75t$]{
			  \includegraphics[width=5.5cm]{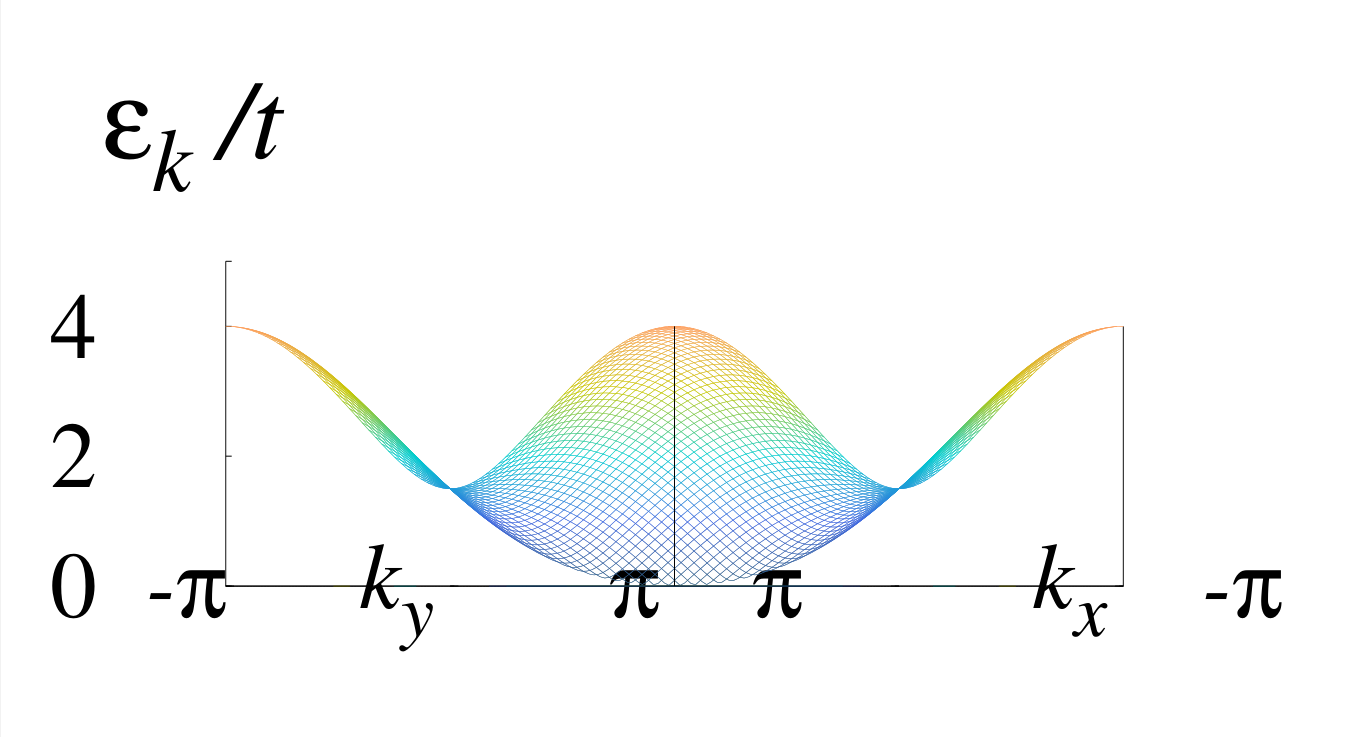}
	}
	\subfloat[$\mu=t$]{
			  \includegraphics[width=5.5cm]{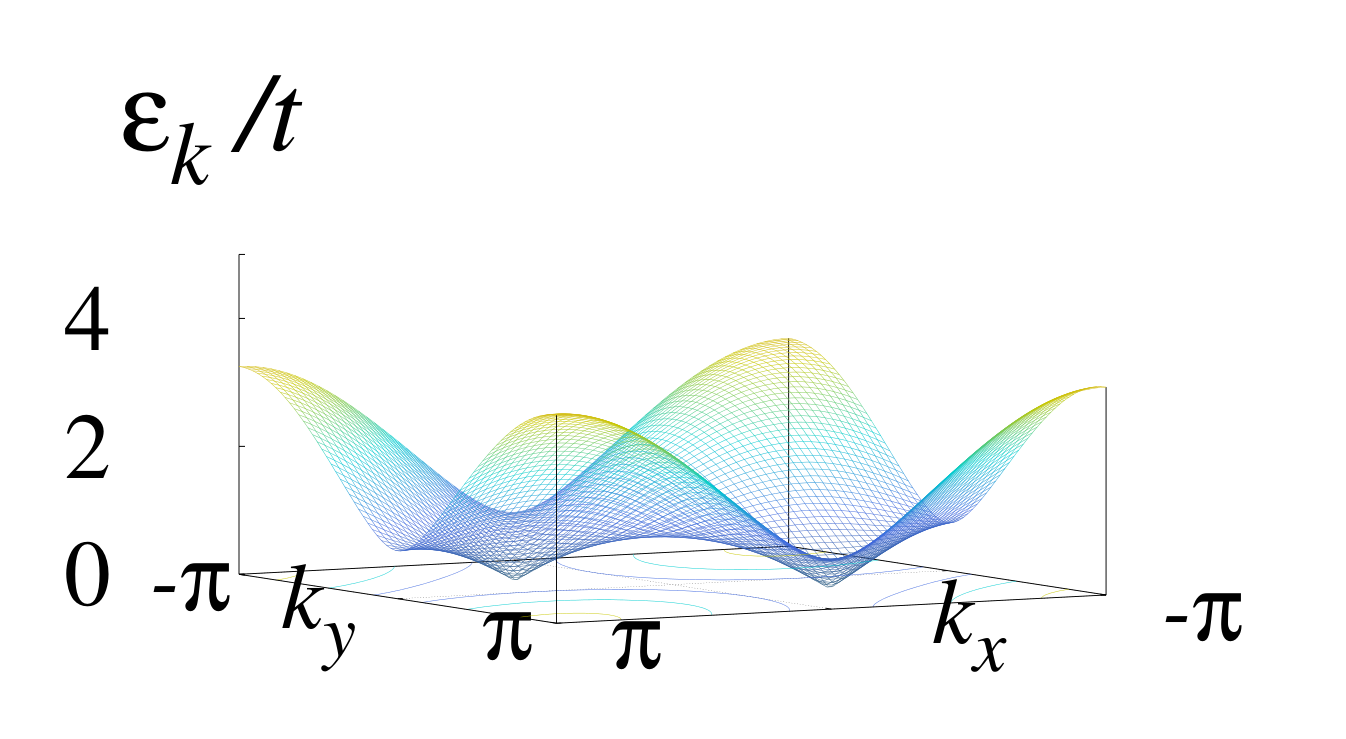}
	}

	\subfloat[$\mu=0.25t$]{
			  \includegraphics[width=5.5cm]{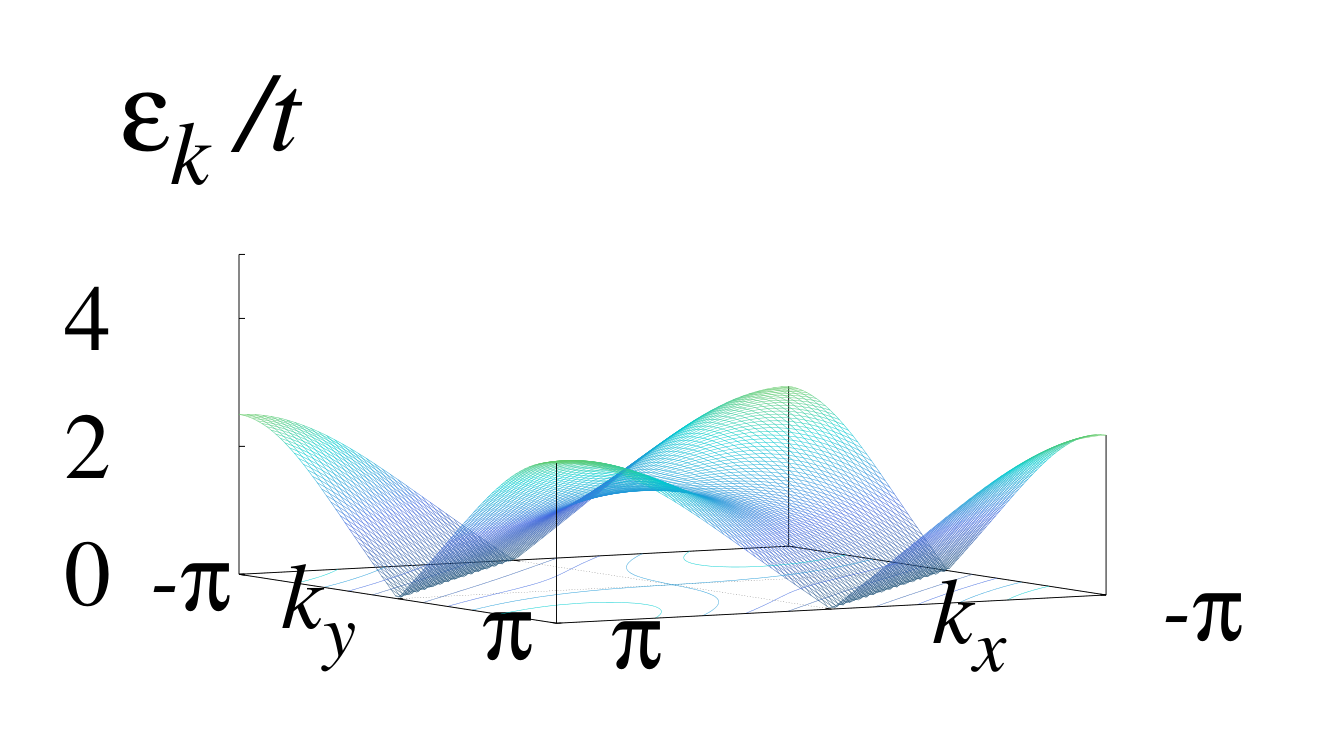}
	}
	\subfloat[$\mu=-t$]{
			  \includegraphics[width=5.5cm]{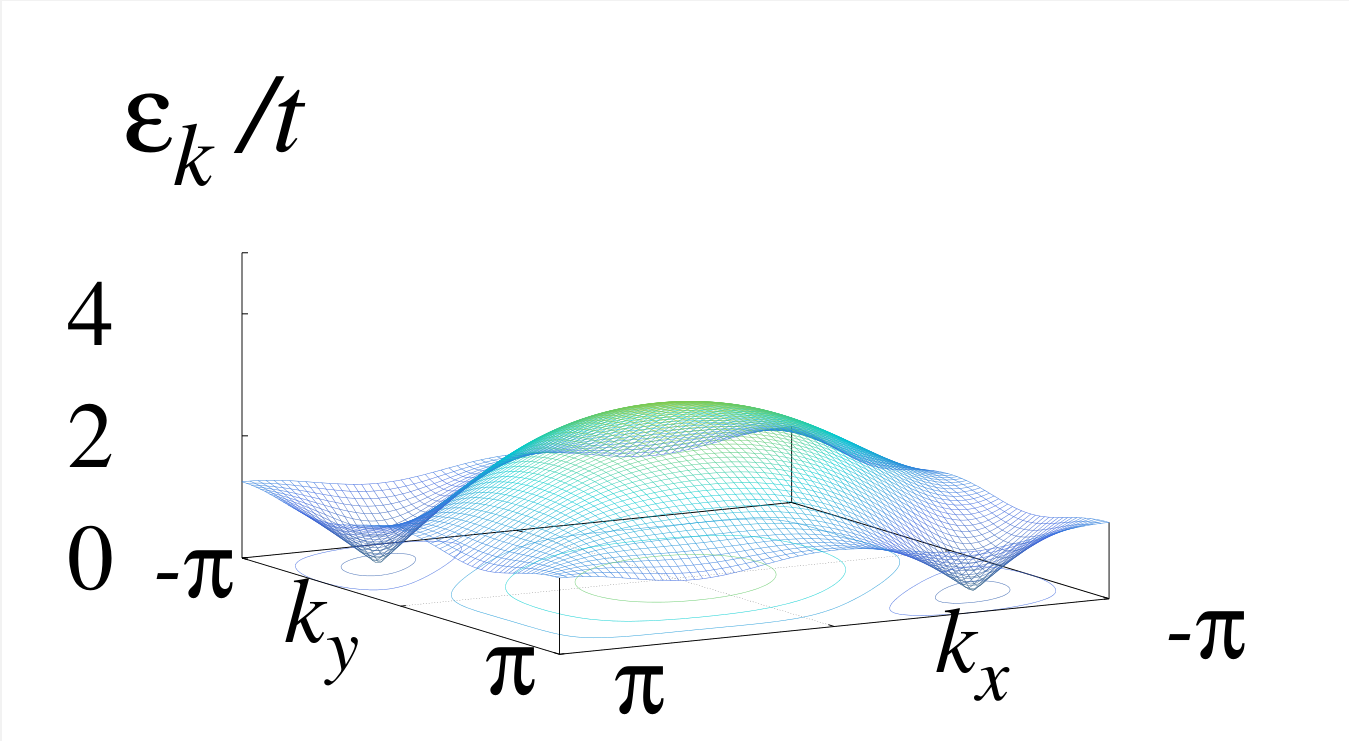}
	}
	\subfloat[$\mu=-2.25t$]{
			  \includegraphics[width=5.5cm]{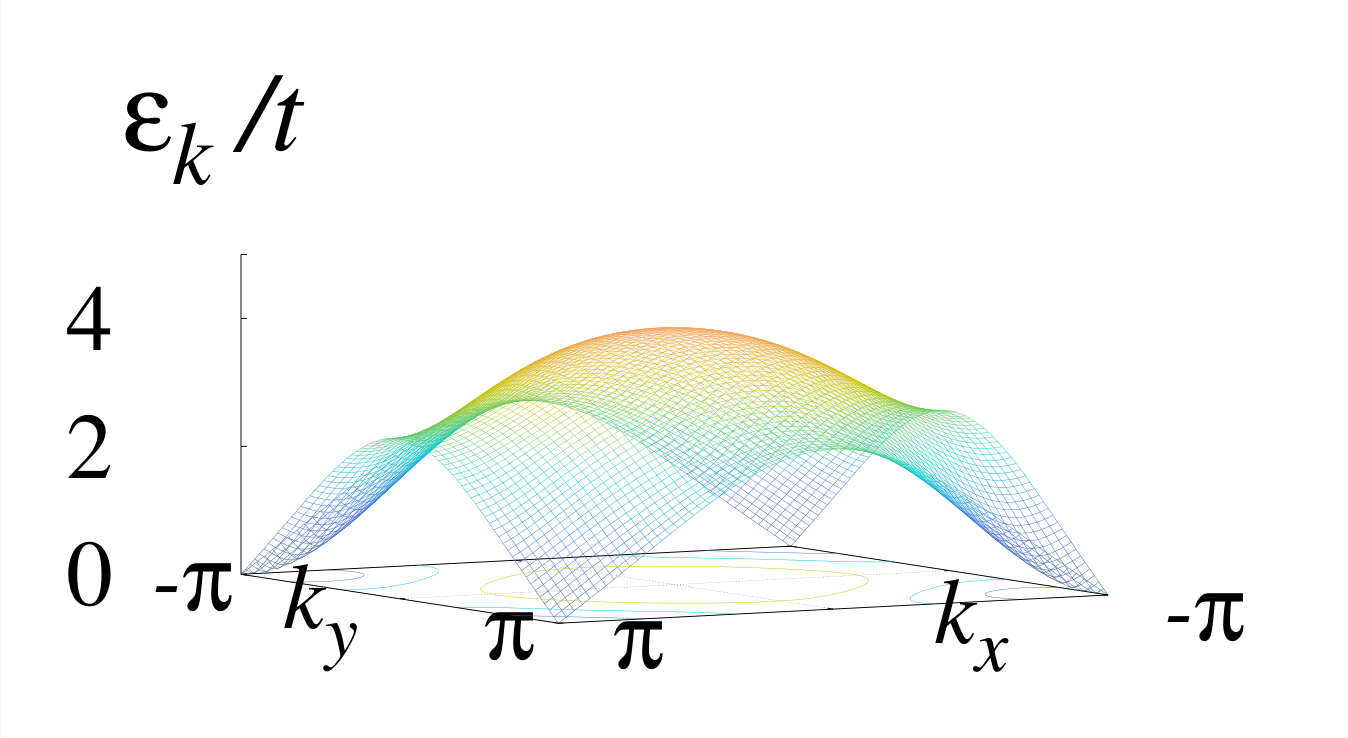}
	}

		\caption{Energy dispersions of the modified square-lattice Kitaev
		model for several values of chemical potential $\mu$ with $\tdp=-0.25t$ and $\Delta=t$. 
					 The energy dispersions correspond to the empty circles on the line $\tdp=-0.25t$ in Fig.~\ref{phase_diagram}.} \label{Kitaev_phase_1}
\end{figure*}
\begin{figure*}

	\subfloat[$\mu=2t$]{
		\includegraphics[width=5.5cm]{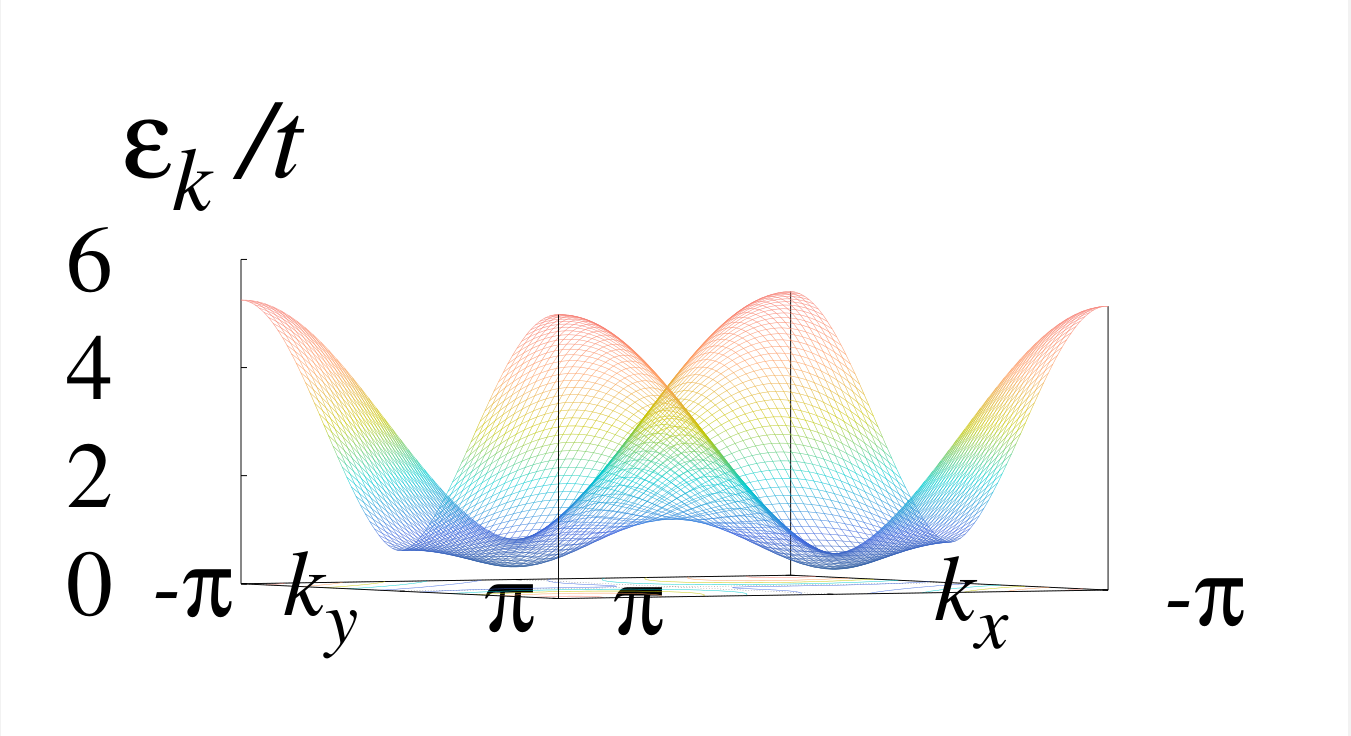}
	}
	\subfloat[$\mu=1.65t$]{
		\includegraphics[width=5.5cm]{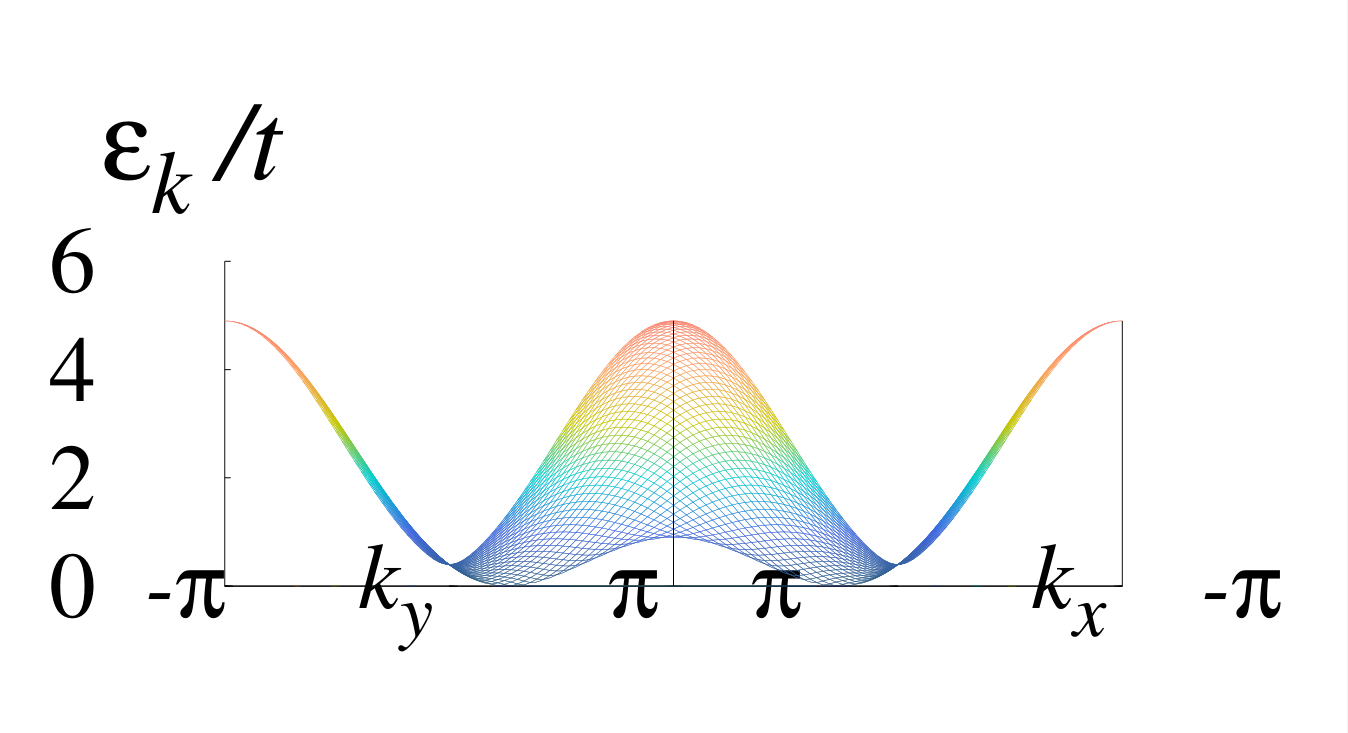}
	}
	\subfloat[$\mu=0.75t$]{
		\includegraphics[width=5.5cm]{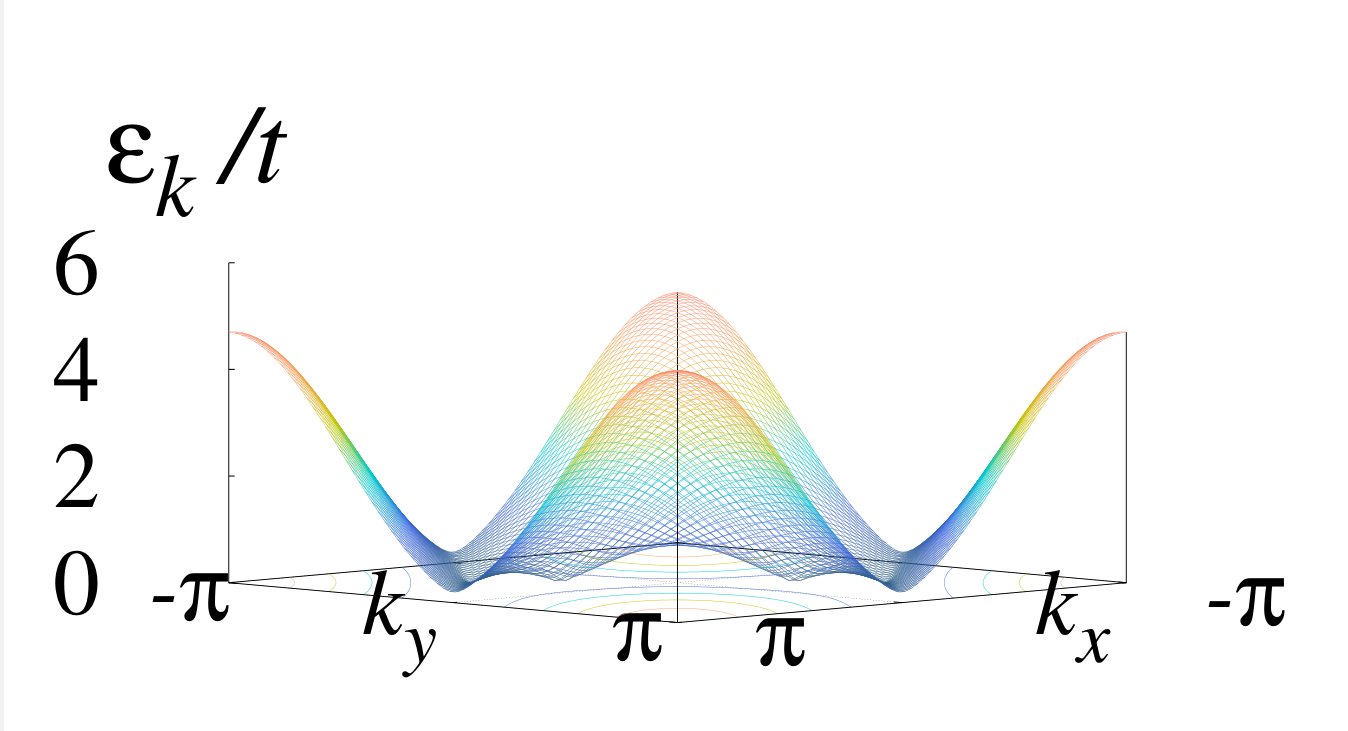}
	}

	\subfloat[$\mu=1.25t$]{
		\includegraphics[width=5.5cm]{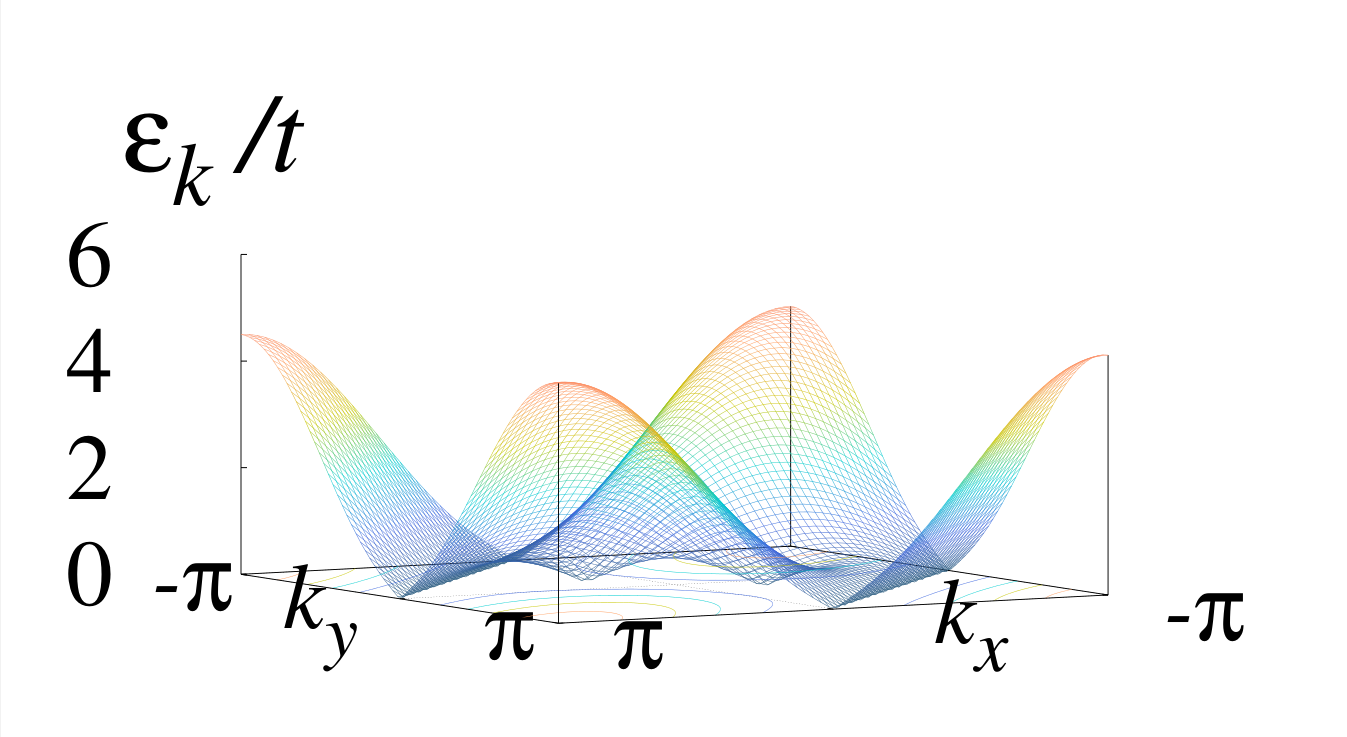}
	}
	\subfloat[$\mu=t$]{
		\includegraphics[width=5.5cm]{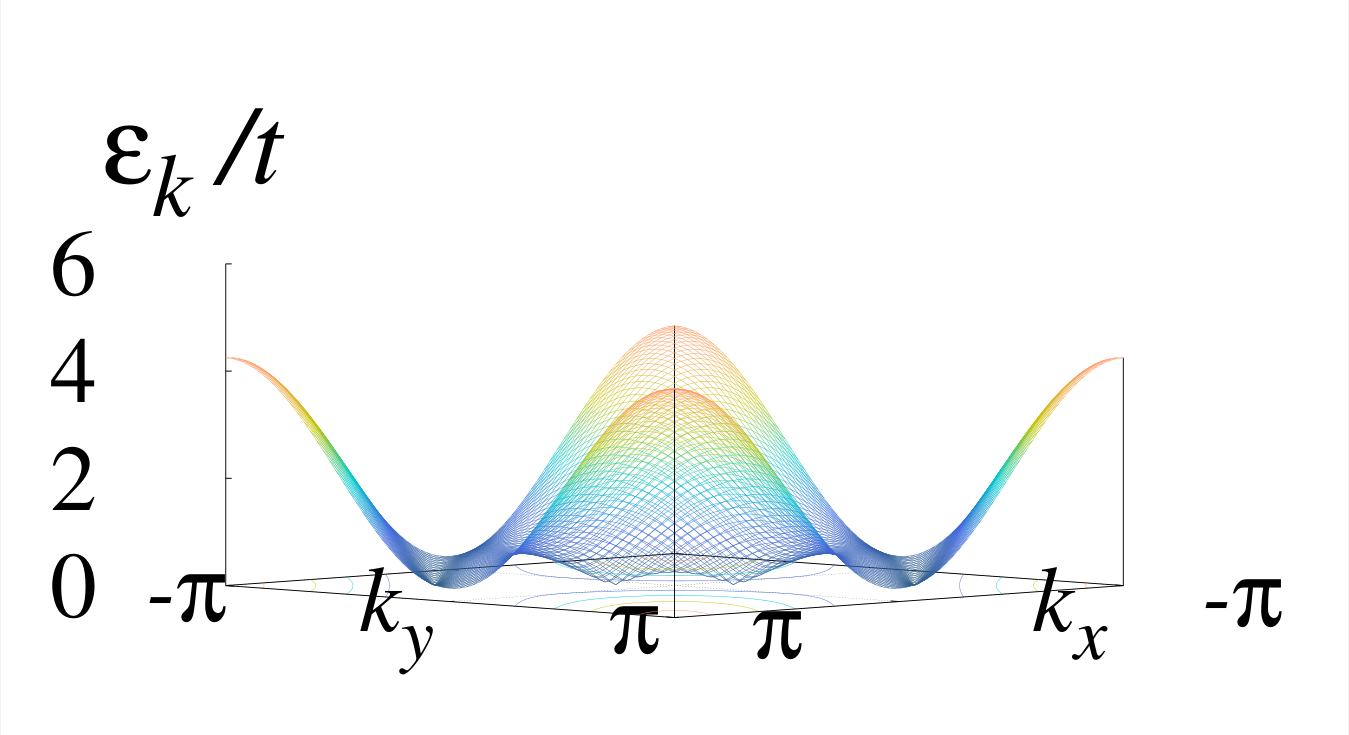}
	}
	\subfloat[$\mu=0.75t$]{
		\includegraphics[width=5.5cm]{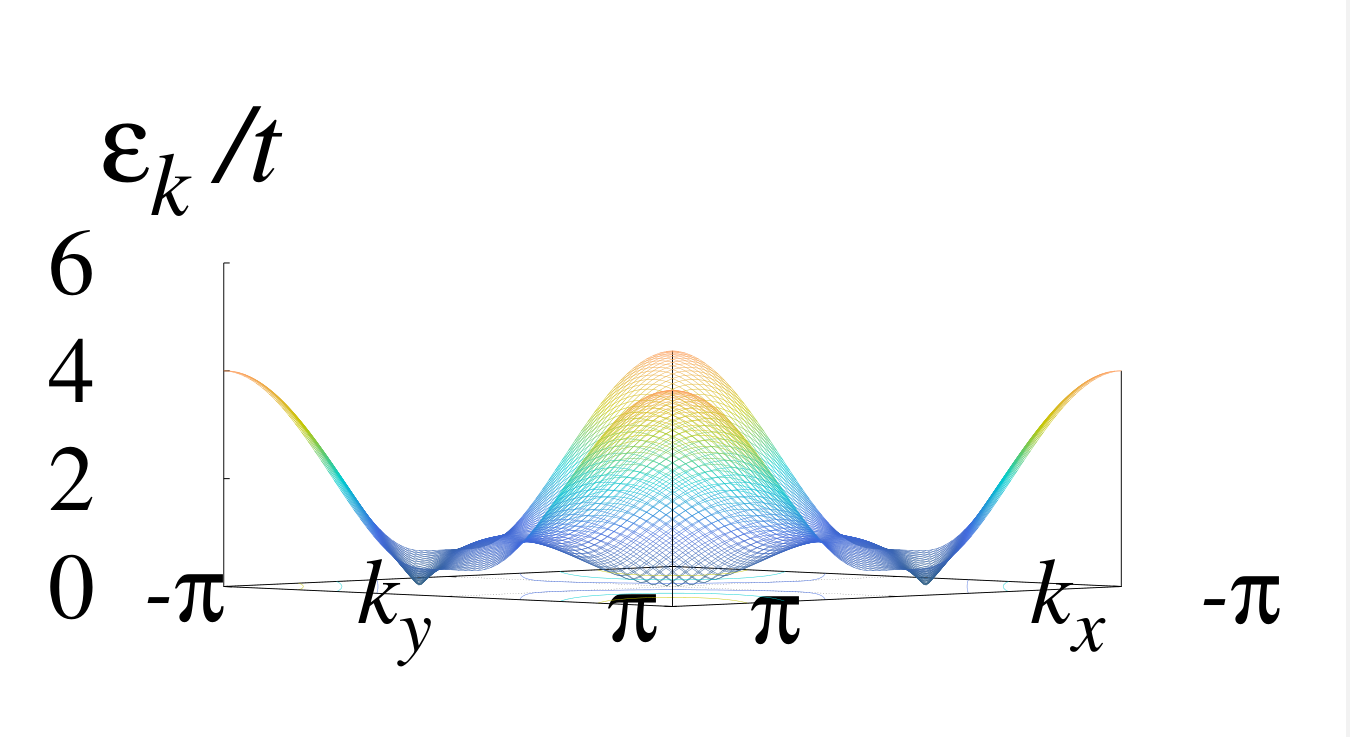}
	}

	\subfloat[$\mu=0$]{
		\includegraphics[width=5.5cm]{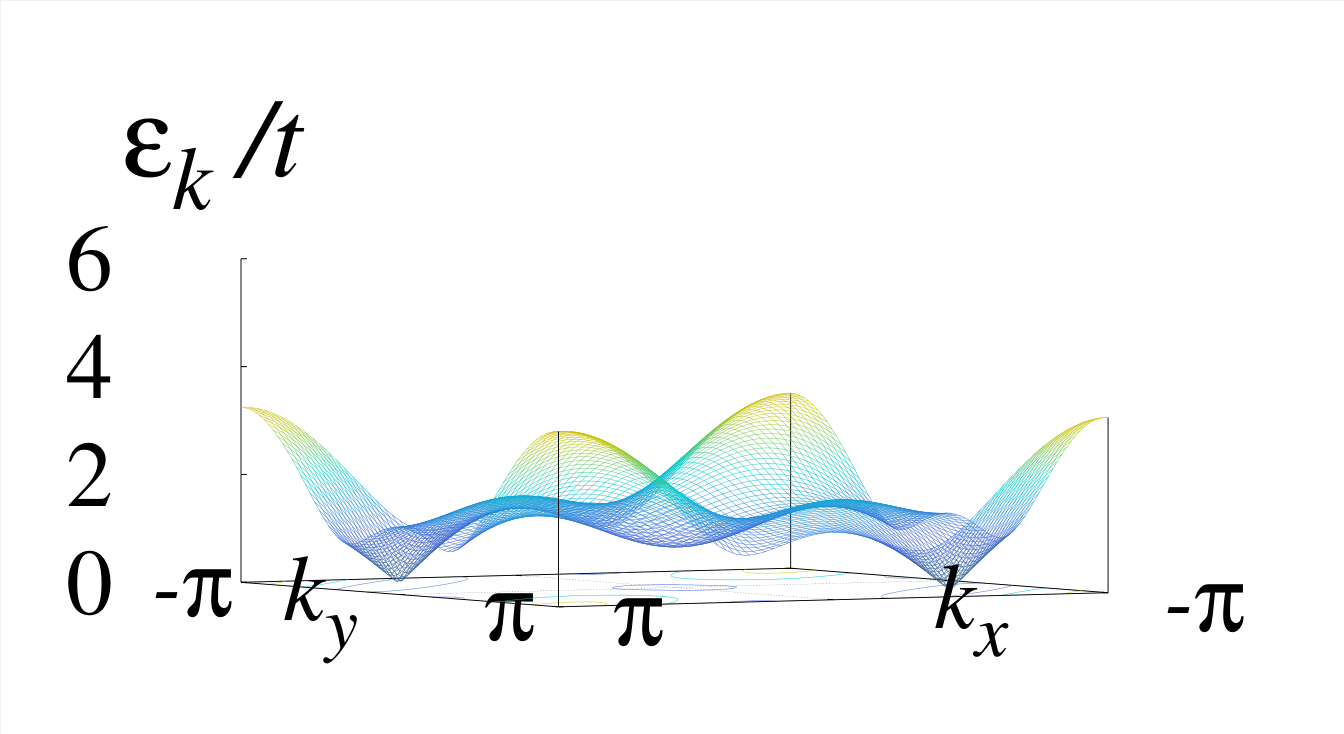}
	}
	\subfloat[$\mu=-3.25t$]{
		\includegraphics[width=5.5cm]{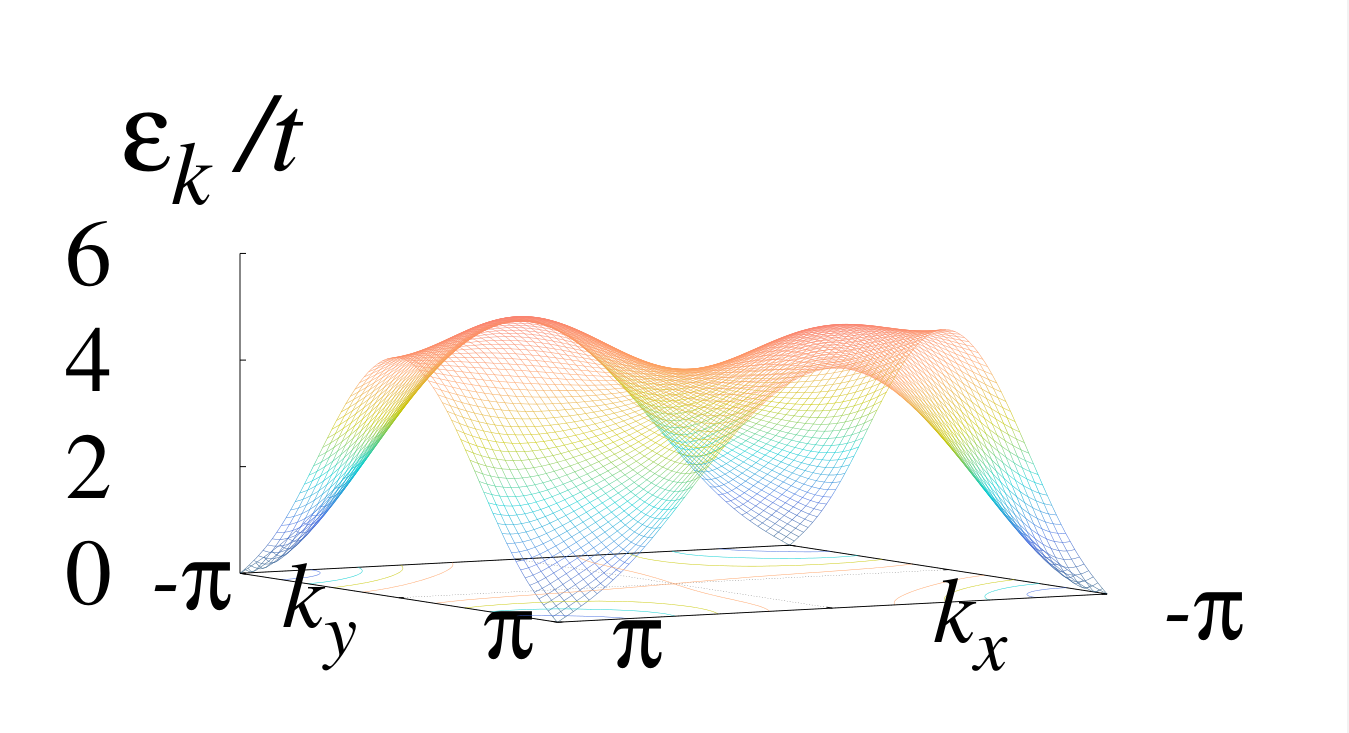}
	}
	\subfloat[$\mu=-4t$]{
		\includegraphics[width=5.5cm]{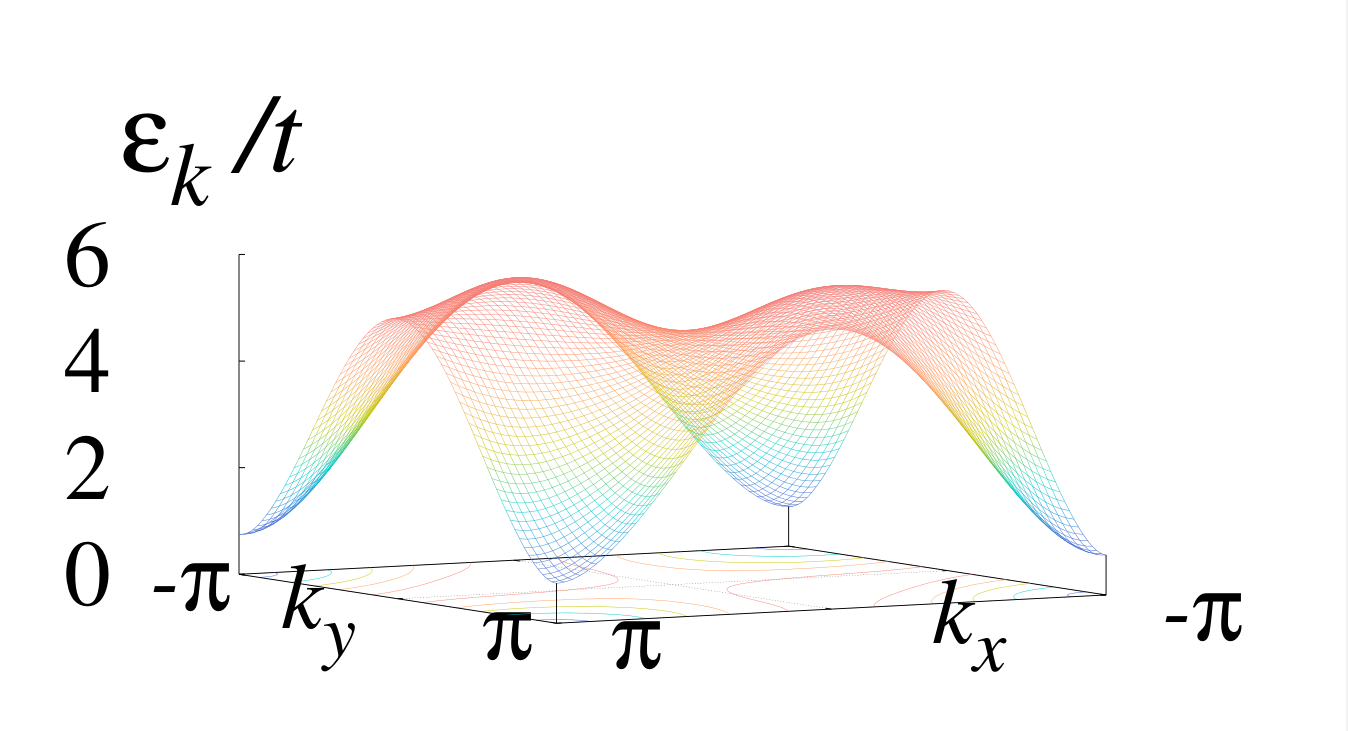}
	}

		\caption{Energy dispersions of the modified square-lattice Kitaev model 
					for several values of chemical potential $\mu$ with $\tdp=-1.25t$ and $\Delta=t$. 
					 The energy dispersions correspond to the filled circles on the line $\tdp=-1.25t$ in Fig.~\ref{phase_diagram}.} \label{Kitaev_phase_2}
\end{figure*}

The evolution of the energy dispersions with the chemical potential $\mu$ can be
understood in terms of how  two quasiparticle bands overlap each other. 
In Fig.~\ref{Kitaev_phase_1} we plot the energy dispersions as $\mu$
decreases for $\tdp=-0.25t$, which illustrates well the general behavior of the energy
dispersions for $-\frac{1}{2}t<\tdp<0$. 
For $\mu > t+\tdp$ the two quasiparticle bands are separated by a full gap
(Fig.~\ref{Kitaev_phase_1}(a)) and they
touch each other at one gapless point $(k_x,k_y)=(0,0)$ when $\mu = t+\tdp$
(Fig.~\ref{Kitaev_phase_1}(b)). 
Further decrease of $\mu$ forces the gapless point
to split into a pair of gapless points (Fig.~\ref{Kitaev_phase_1}(c)), 
which are located on the line $k_{y}=-k_{x}$ 
since the pairing terms in the Hamiltonian impose the condition for zero energy.
The existence of the pair is
robust against the change of the chemical potential $\mu$ in that phase, implying their topological origin.
Such a pair persists until they merge
into a single point at $(k_x,k_y)=(\pi,-\pi)$ (Fig.~\ref{Kitaev_phase_2}(f)). 
It is of interest to note that 
a nodal line $k_{y}=k_{x} + \pi$ shows up at $\mu=-\tdp$ 
(Fig.~\ref{Kitaev_phase_2}(d)) while the two gapless points approach the merging
point $(\pi,-\pi)$. 
The band gap reopens for $\mu<-2t+\tdp$. 

For $\tdp<-\frac{1}{2}t$, we can observe qualitatively different behavior of
the energy dispersions, which are demonstrated in Fig.~\ref{Kitaev_phase_2}
in the case of $\tdp=-1.25t$.
When $\tdp<-\frac{1}{2}t$, $\epsilon_{\bm{k}}$ has
two minima before the two band touch each other (Fig.~\ref{Kitaev_phase_2}(a)), which is different from the former case. 
When the band gap closes at these points, two
gapless points are formed (Fig. \ref{Kitaev_phase_2}(b)). 
Each gapless point splits into two gapless points with further reduction of the
chemical potential and they move away from each other along $k_{y}=-k_{x}$ (Fig. \ref{Kitaev_phase_2}(c)). 
At $\mu=-\tdp$
a nodal line $k_{y}=k_{x}+\pi$ also appears as in the case of $-\frac12 t<\tdp<0$
while 
the inner pair of gapless points still survive (Fig.~\ref{Kitaev_phase_2}(d)). 
The inner pair merges at $(k_x,k_y)=(0,0)$ when $\mu=2t+\tdp$ and only one pair of
gapless points remains in the system
(Fig.~\ref{Kitaev_phase_2}(f)).
Finally 
the outer pair merges at $(k_x,k_y)=(\pi,-\pi)$ when $\mu=-2t-\tdp$ (Fig.~\ref{Kitaev_phase_2}(h))
and a finite gap reopens for $\mu < -2t-\tdp$ (Fig.~\ref{Kitaev_phase_2}(i)). 

\begin{figure*}
		\includegraphics[width=6.65cm]{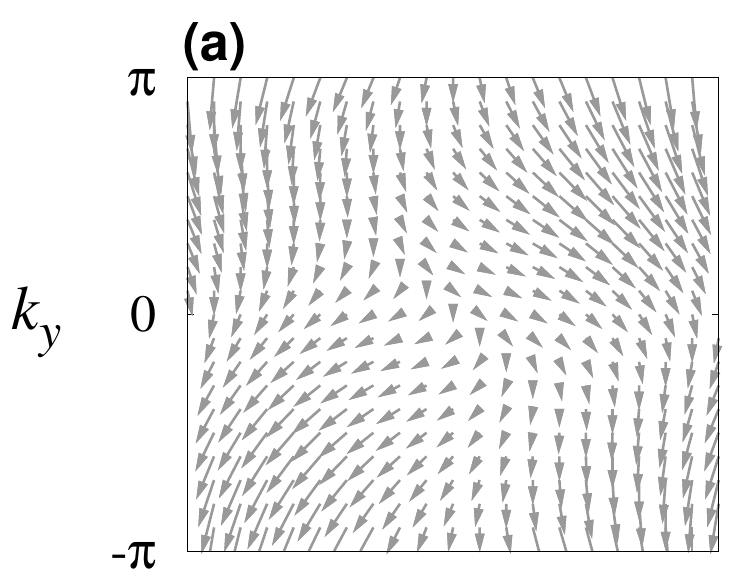}
		\includegraphics[width=5.0cm]{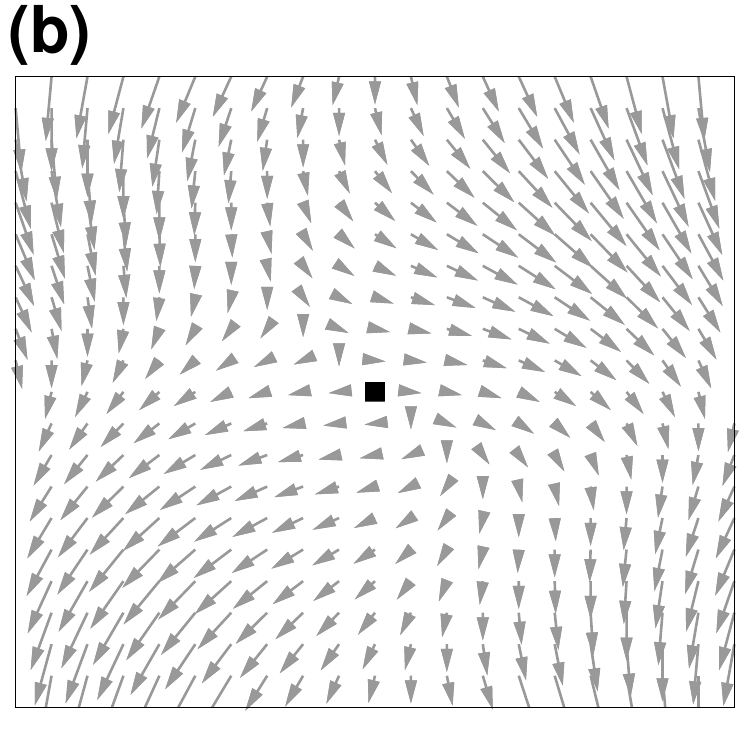}
		\includegraphics[width=5.0cm]{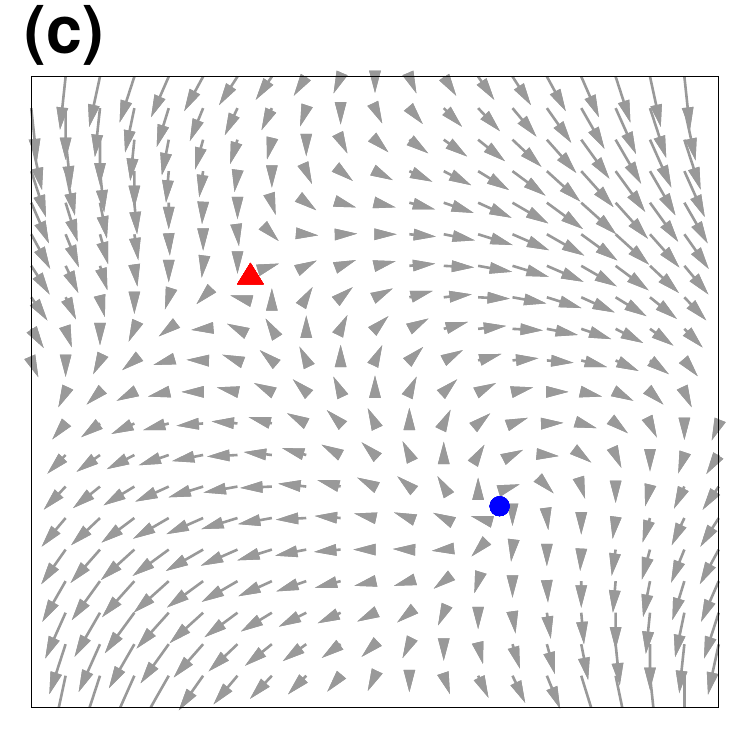}

		\includegraphics[width=6.65cm]{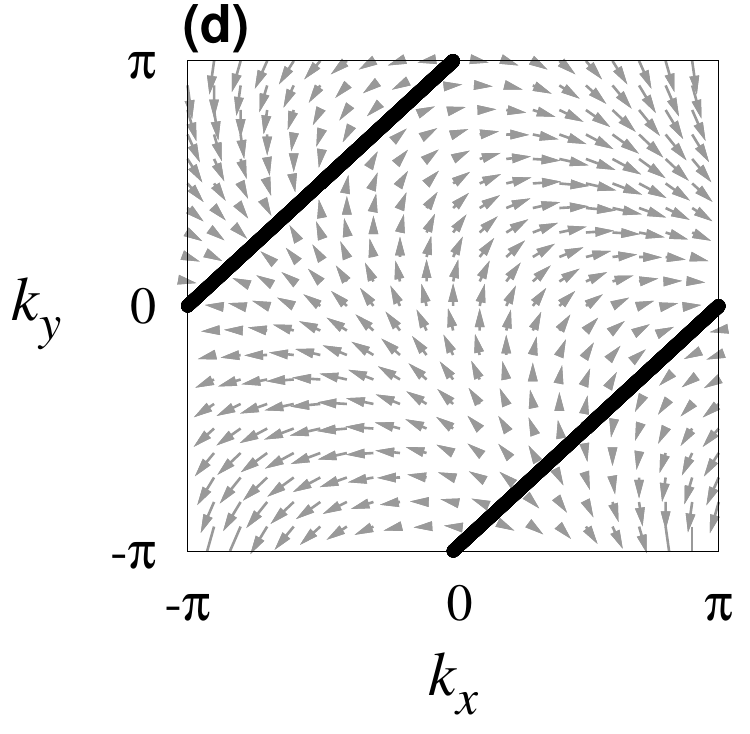}
		\includegraphics[width=5.0cm]{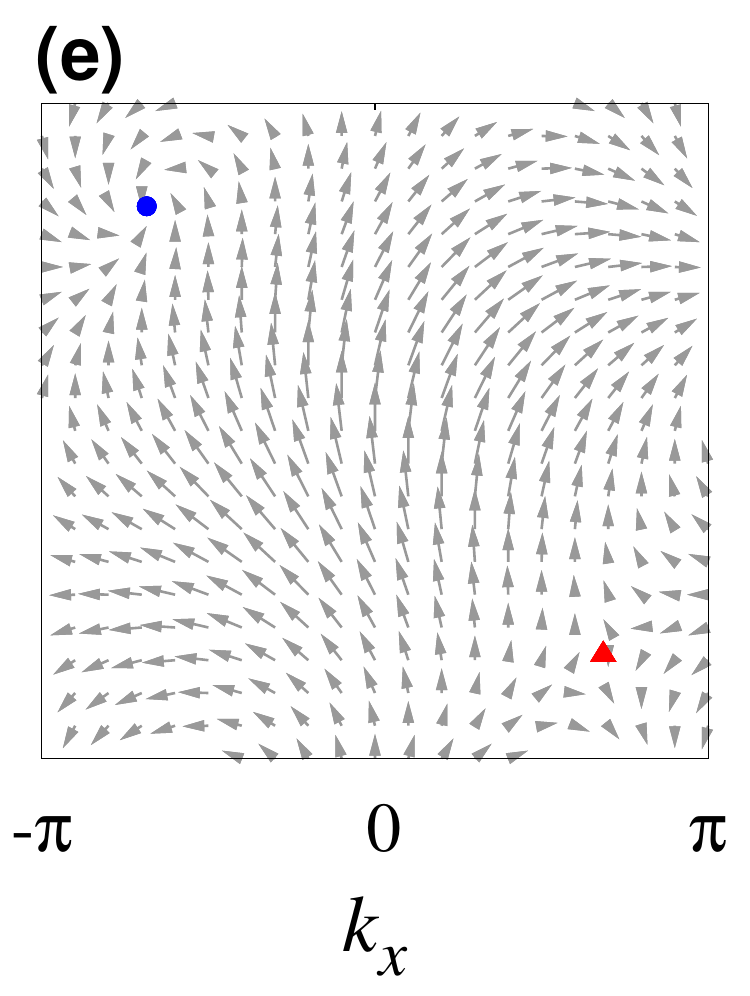}
		\includegraphics[width=5.0cm]{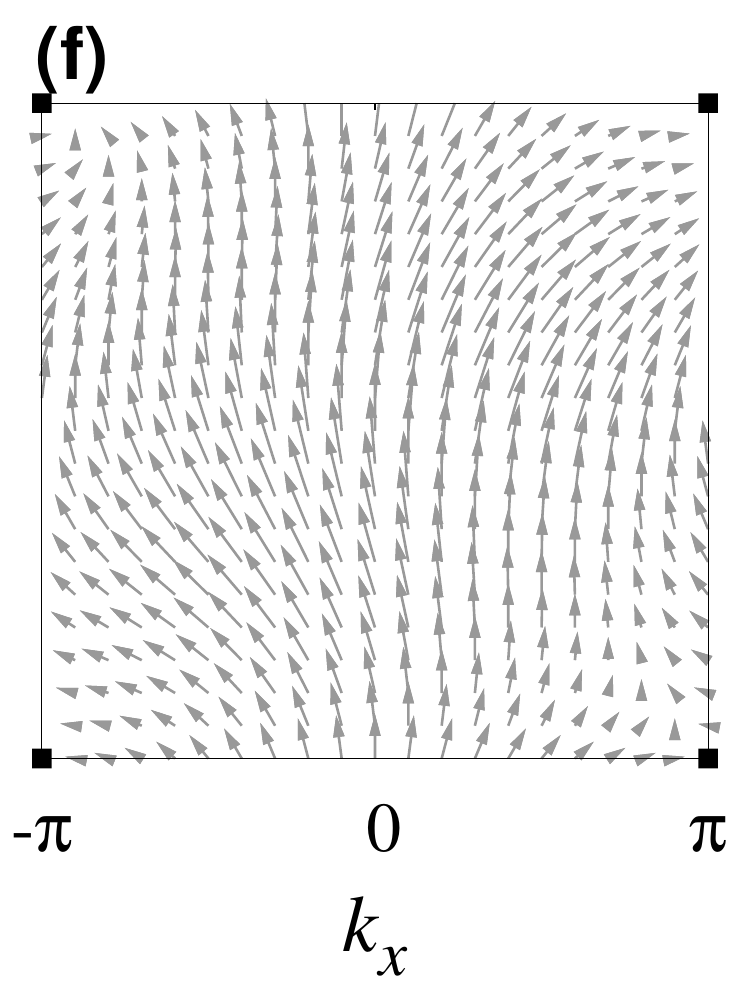}

	
		  \caption{ The vector fields $\bm{B}(\bm{k})$ of the modified square-lattice Kitaev
		  model with $\tdp=-0.25t$ and $\Delta=t$
		  for $\mu=$ (a) $2t$, (b) $1.75t$, (c) $t$, (d) $0.25t$, (e) $-t$, and (f) $-2.25t$, 
		  which correspond to the empty
		  circles on the line $\tdp=-0.25t$ in Fig.~\ref{phase_diagram}.
		  The horizontal and vertical components of the arrows at momentum
		  $\bm{k}$
		  correspond to $B_{1}(\bm{k})$ and $B_{2}(\bm{k})$, respectively.
		  Red triangles, blue circles
		  and black squares represent the gapless points of 
		  a winding number -1, 1, and 0, respectively. 
		  }
		  \label{Kitaev_phase_3} 
\end{figure*}

\subsection{Topological invariants of gapless points}
\begin{figure*}
						
		\includegraphics[width=6.65cm]{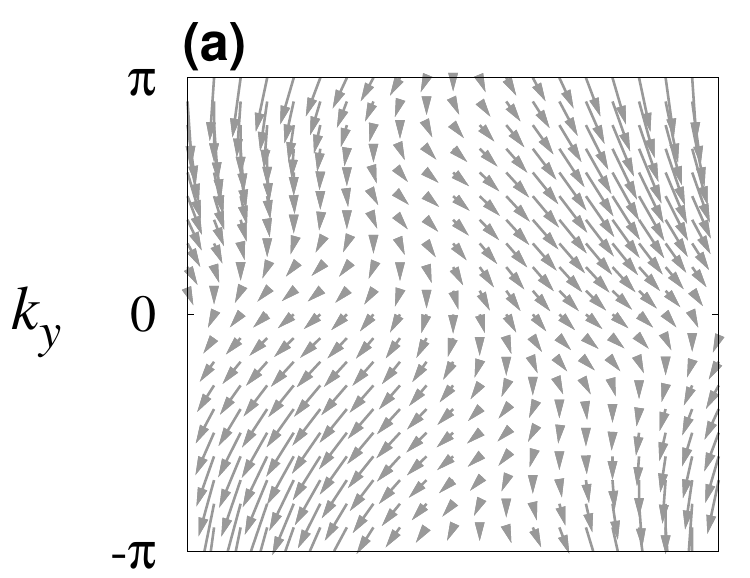}
		\includegraphics[width=5.0cm]{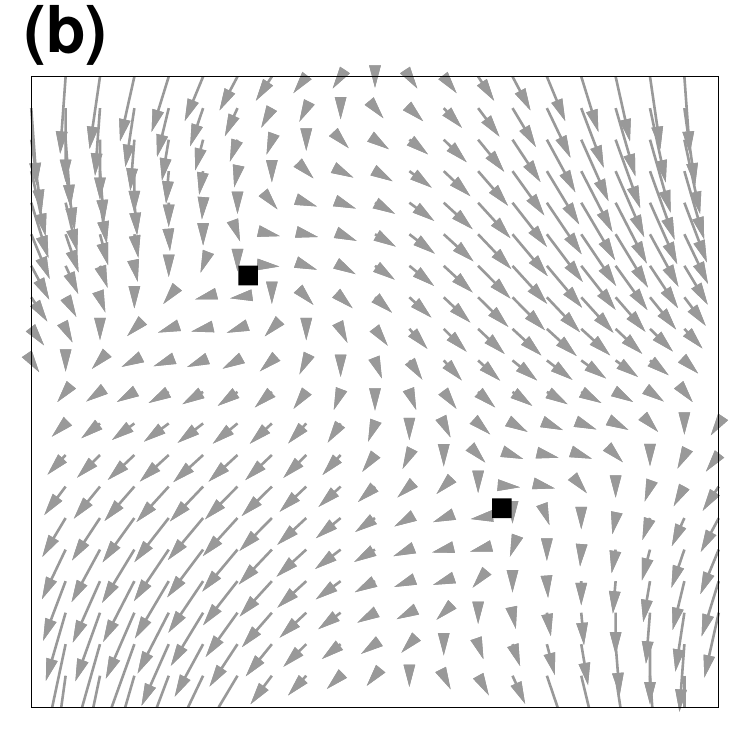}
		\includegraphics[width=5.0cm]{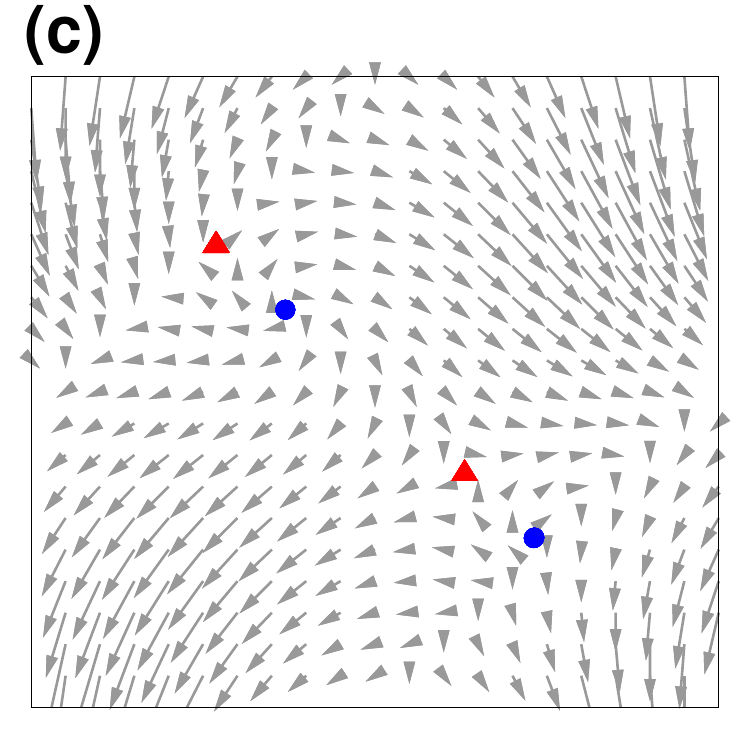}

		\includegraphics[width=6.55cm]{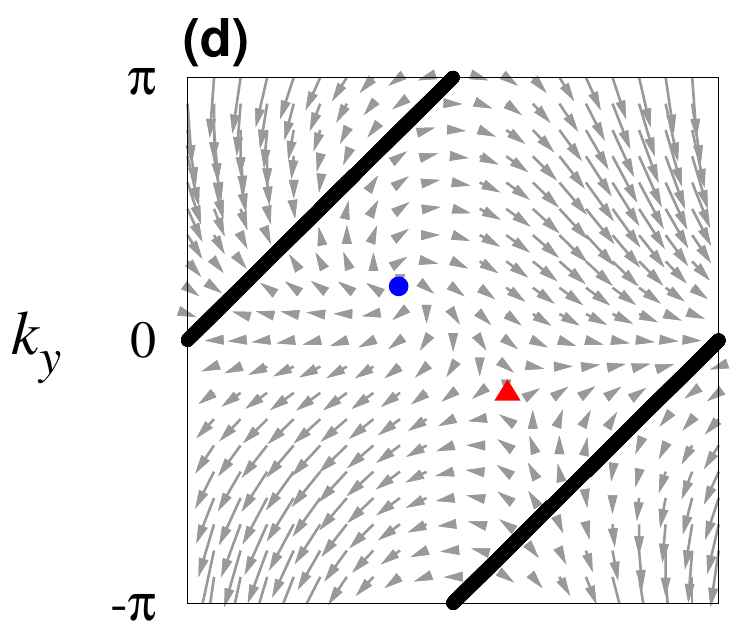}
		\includegraphics[width=5.0cm]{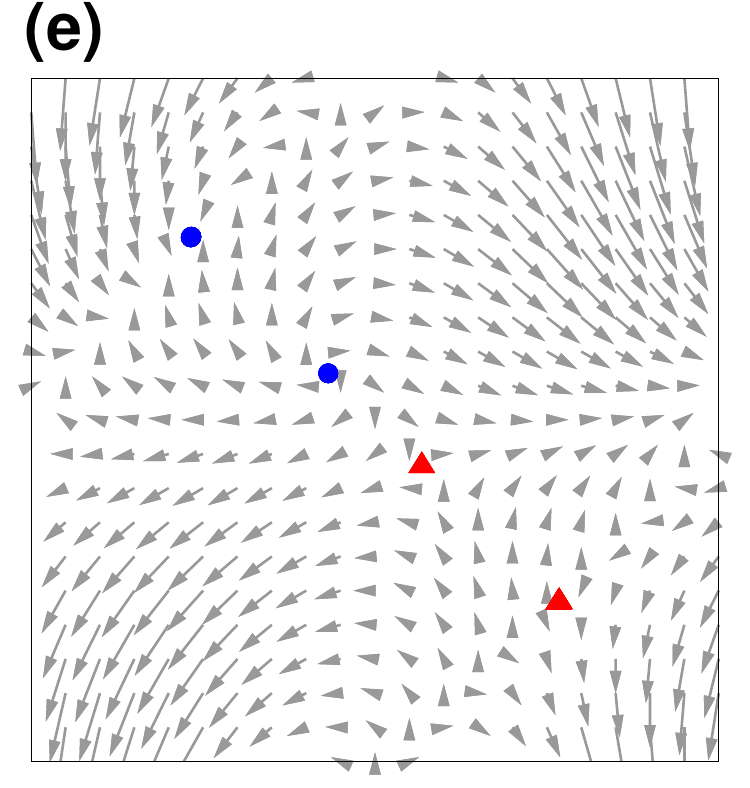}
		\includegraphics[width=5.0cm]{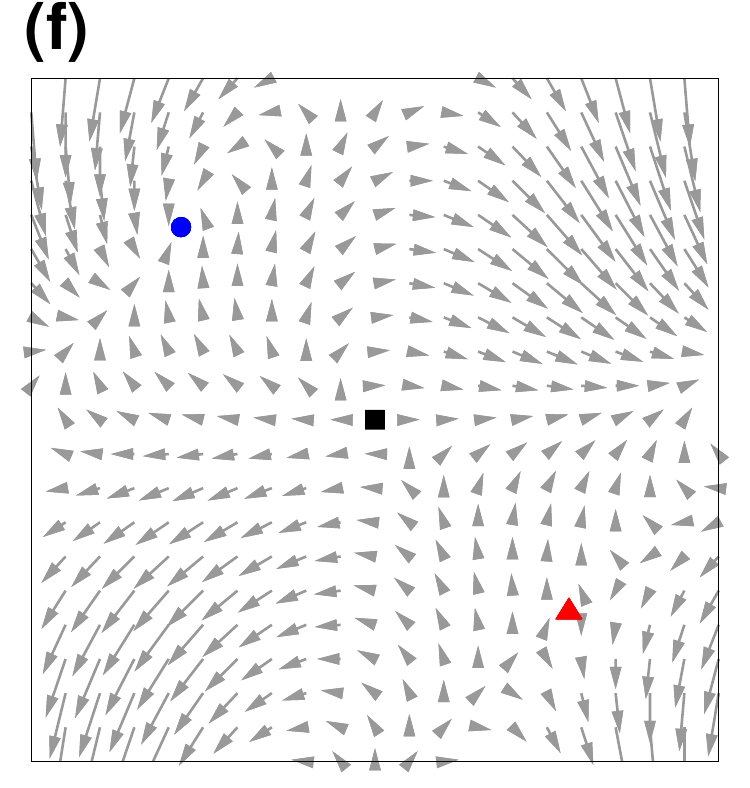}

		\includegraphics[width=6.65cm]{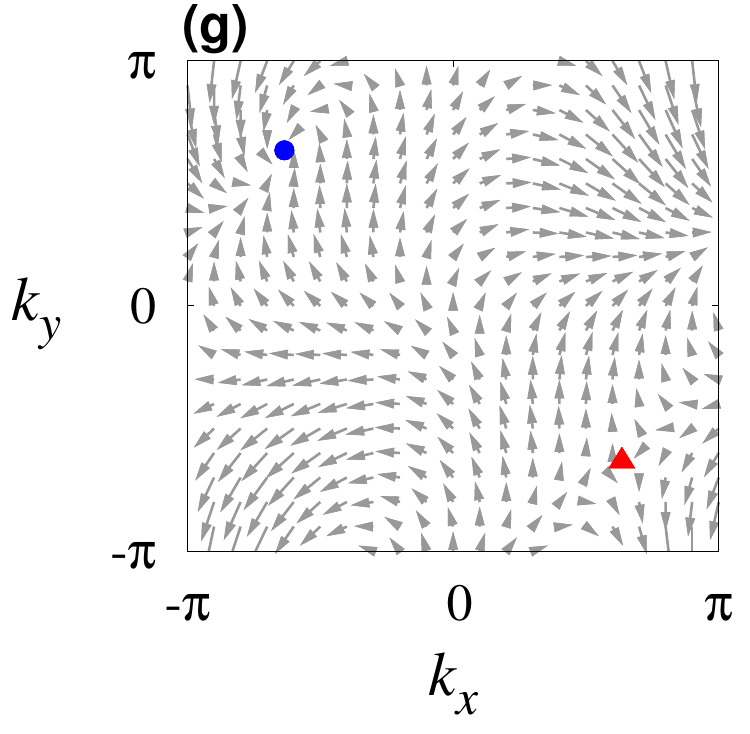}
		\includegraphics[width=5.0cm]{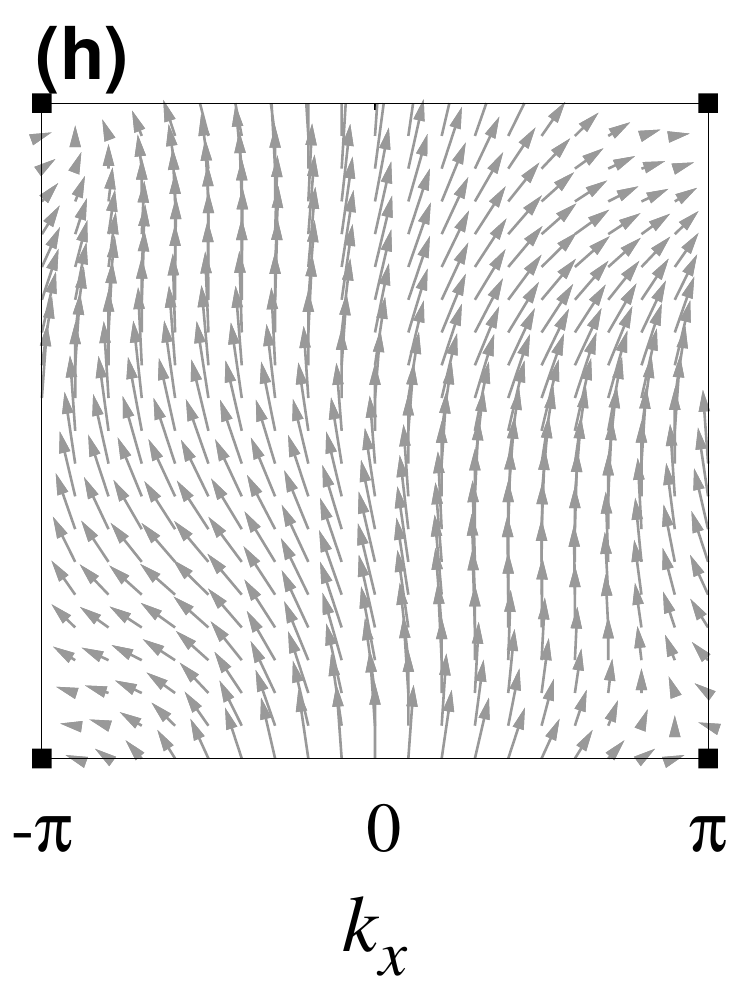}
		\includegraphics[width=5.0cm]{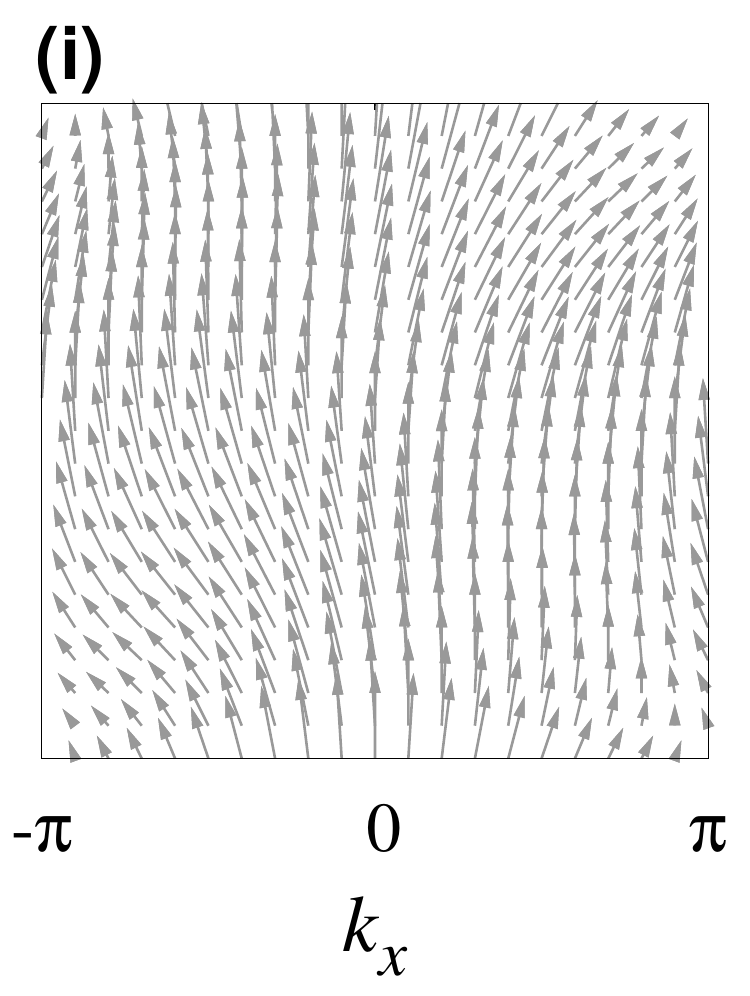}

 		\caption{ The vector fields $\bm{B}(\bm{k})$ of the modified square-lattice Kitaev
		  model with $\tdp=-1.25t$ and $\Delta=t$
		  for $\mu=$ (a) $2t$, (b) $1.65t$, (c) $1.45t$, (d) $1.25t$,
		  (e) $t$, (f) $0.75t$, (g) $0$, (h) $-3.25t$, and (i) $-4t$,
		  which correspond to the filled
		  circles on the line $\tdp=-1.25t$ in Fig.~\ref{phase_diagram}.
		  The gapless points are marked by the same symbols as in
		  Fig.~\ref{Kitaev_phase_3}.
		  }
		  \label{Kitaev_phase_4} 
\end{figure*}

We next examine the topological characters of individual gapless points
and 
express the Hamiltonian matrix $h_{\bm{k}}$ 
\begin{equation}
		  h_{\bm{k}}=\sum_{i=1}^{2} B_{i}(\bm{k})\sigma_{i}
\end{equation}
in terms of the vector field 
 $\bm{B}(\bm{k})=(B_{1},\,B_{2})$ 
\begin{align}
	B_{1}(\bm{k})&=\Delta\left(\sin{k_{x}}+\sin{k_{y}}\right),
	\nonumber
	\\
	B_{2}(\bm{k})&=-\mu+t\left(\cos{k_{x}}+\cos{k_{y}}\right)+\tdp\cos(k_{x}-k_{y}),
\end{align}
and Pauli matrices
\begin{align}
	\sigma_1 =\left(
		\begin{matrix}
			0 & -i
			\\
			i & 0
		\end{matrix}
	\right),
	\quad \quad
	\sigma_2 =\left(
		\begin{matrix}
			1 & 0
			\\
			0 & -1
		\end{matrix}
	\right).
\end{align}
The gapless points exist in the form of vortices (see Figs.~\ref{Kitaev_phase_3}
and \ref{Kitaev_phase_4}).
The winding number $\omega$ of each gapless point can be calculated conveniently 
by the expansion around the vortex center $\bm{k}_{0}$, 
where $\bm{k}_{0}$ is the location of the topological defect. 
Around the vortex center
the vector field $\bm{B}(\bm{k})$ is expanded up to a linear order in $\bm{q}
\equiv \bm{k}-\bm{k}_0$:
\begin{align}
	\nonumber
	B_{1}(\bm{q})=&\Delta(q_{x}\cos{k_{0x}} +q_{y}\cos{k_{0y}} ),
	\\
	B_{2}(\bm{q})=&[-t\sin{k_{0x}}-\tdp\sin(k_{0x}-k_{0y})]q_{x}
	\nonumber
	\\
	& + [-t\sin{k_{0y}}+\tdp\sin(k_{0x}-k_{0y})]q_{y}.
	\label{av_field}
\end{align}
The matrix $h_{\bm{k}}$ is then given  in the form
\begin{align}
	h_{\bm{k}}=\sum_{i,j=1}^{2}a_{ij}q_{i}\sigma_{j}
\end{align}
with $(q_1,q_2) \equiv (q_x,q_y)$, leading to  
the winding number $\omega$ of the individual topological gapless point
\begin{eqnarray}
	\omega&=&\textrm{sgn}[\textrm{det}(\bm{a})]
	\nonumber
	\\&=&\textrm{sgn}[
			  \Delta\sin(k_{0x}-k_{0y})\left\{t+\tdp(\cos{k_{0x}}+\cos{k_{0y}})\right\}
	].
	\nonumber
			  \\
\end{eqnarray}

We first consider the case of $-\frac{1}{2}t<\tdp<0$. 
In Fig.~\ref{Kitaev_phase_3} we plot the vector fields $\bm{B}(\bm{k})$ 
in the modified square-lattice Kitaev model with $\tdp=-0.25t$
for several values of chemical potential $\mu$.
In the system with $\mu=t+\tdp$ the band gap closes, yielding a gapless point of zero winding
number at $\bm{k}=(0,0)$ (see Fig.~\ref{Kitaev_phase_3}(b)). 
When the chemical potential becomes lower than this value,
the gapless point splits into
two Weyl-type gapless points with linear energy dispersion, 
one with the winding number -1 and the other with 1,
while total winding number is conserved (Fig.~\ref{Kitaev_phase_3}(c)). 
The winding number of each Weyl point is not changed by
the continuous change of the chemical potential $\mu$. 
At $\mu=-\tdp$, a nodal line appear at $k_y=k_x+\pi$,
making significant changes in the configuration of the vector field $\bm{B}(\bm{k})$ 
(Fig.~\ref{Kitaev_phase_2}(d)). 
For the chemical potential below $-\tdp$, the winding
numbers of the gapless points are interchanged (Fig.~\ref{Kitaev_phase_2}(e)). 
The two topological defects which have opposite winding numbers
merge into one with zero winding number at ($\pi,-\pi$) for 
$\mu=-2t+\tdp$ (Fig.~\ref{Kitaev_phase_3}(f)).
Below this chemical potential the systems exhibits no gapless points with a
finite band gap.

\begin{figure}
		  \subfloat[]{
		  	\includegraphics[width=7cm]{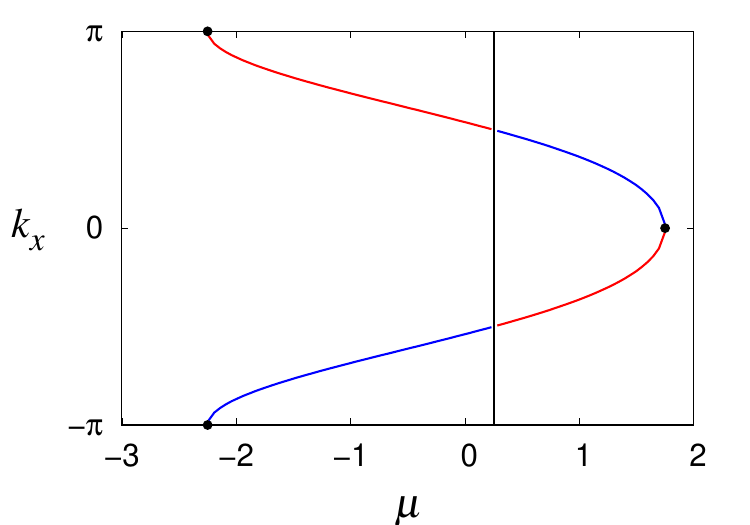}
 }
 \\
		  \subfloat[]{
	\includegraphics[width=7cm]{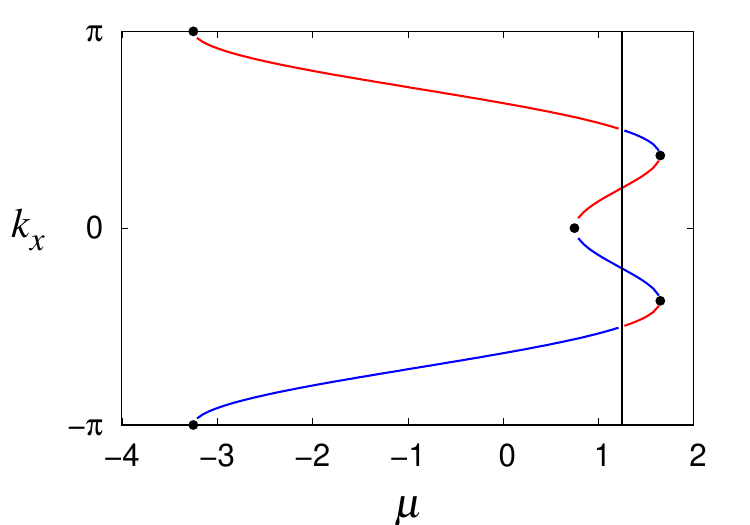}
	}
		  \caption{ The $x$-component of momentum, $k_{x}$, of the topological defects as a
		  function of the chemical potential $\mu$ for (a) $\tdp=-0.25t$ and (b)
		  $\tdp=-1.25t$. 
		  Topological defects of winding number $-1/0/1$ are marked by red/black/blue lines.} 
		  \label{loc}

\end{figure}

The general features of the vector field $\bm{B}(\bm{k})$ for $\tdp<-\frac{1}{2}t$
are demonstrated well in Fig.~\ref{Kitaev_phase_4}, which 
shows the vector fields $\bm{B}(\bm{k})$ for $\tdp=-1.25t$.
In contrast to the case with $-\frac{1}{2}t<\tdp<0$,
the system exhibits two gapless points when the band gap closes;
the winding numbers of both gapless points are zero (Fig.~\ref{Kitaev_phase_4}(b)). 
Each topological defect splits into two gapless points, resulting in two
pairs of Weyl points whose winding numbers are opposite to each other 
(Fig.~\ref{Kitaev_phase_4}(c)). 
As in the case with $-\frac{1}{2}t<\tdp<0$,
a nodal line, $k_y=k_x+\pi$, shows up for $\mu=-\tdp$, 
which interchanges their winding numbers of an outer pair of Weyl points only 
(Fig.~\ref{Kitaev_phase_4}(d)). 
The inner pair of Weyl points approach each other (see
Fig. \ref{Kitaev_phase_4}(e)) and merge at $\bm{k}=(0,0)$ (Fig.~\ref{Kitaev_phase_4}(f)). 
The remaining pair of Weyl points also merge into a trivial gapless point at
$(\pi,-\pi)$ (Fig.~\ref{Kitaev_phase_4}(h)) at a lower chemical potential,
below which
the system lies in a trivial gapful phase without any topological gapless point (Fig.~\ref{Kitaev_phase_4}(i)). 

Figure~\ref{loc} displays the variation of the momentum of gapless points and their
winding numbers with the chemical potential $\mu$ for   $\tdp=-0.25t$ and  $\tdp=-1.25t$.
It is sufficient to plot the $x$-component only of the momentum of gapless points
for the accurate description of the gapless-point position 
since all the isolated topological defects with winding number $\pm 1$ 
are on the line $k_y = - k_x$ and a nodal
line of winding number zero is given by $k_y=k_x+\pi$ in the whole range of
$k_x$.
For $-\frac12 t<\tdp<0$, as shown in Fig.~\ref{loc}(a), 
a gapless point of winding number zero is formed at $\bm{k}=(0,0)$ for
$\mu=2t+\tdp$.
As the chemical potential is reduced , it evolves to two separate topological
Weyl points; the $x$-component momentum of the one with winding number $+1$
increase while the other with winding number $-1$ moves towards $(-\pi,\pi)$.
The nodal line , which shows up for $\mu=-\tdp$, flips the winding number of
both Weyl points.
They approach $(-\pi,\pi)$ monotonically from the opposite directions and
transforms to one gapless point with zero winding number, which disappears for
$\mu<-2t+\tdp$.

For $\tdp<-\frac12 t$, on the other hand, the system exhibits two trivial gapless points
for $\mu=-t^2/(2\tdp)-\tdp$.
From each trivial gapless point two topological Weyl points are produced;
the inner pair of Weyl points, which are closer to the origin, approach each
other, and merge at the origin for $\mu=2t+\tdp$.
The outer pair moves away from the origin and finally meets to form a trivial
gapless point at $(-\pi,\pi)$ for $\mu=-2t+\tdp$. 
Winding numbers are affected by the nodal line, which is formed for $\mu=-\tdp$,
only for the outer pair of Weyl points.

\subsection{Majorana flat band edge modes in ribbon geometry}

In this section, we consider the modified
square-lattice Kitaev model in the ribbon geometry to
examine the topological edge states which is characteristics of topological
bulk states in the system.
It turns out that the accompanying edge states are Majorana flat bands formed at
the edges.

\begin{figure*}
		\includegraphics[width=6.86cm]{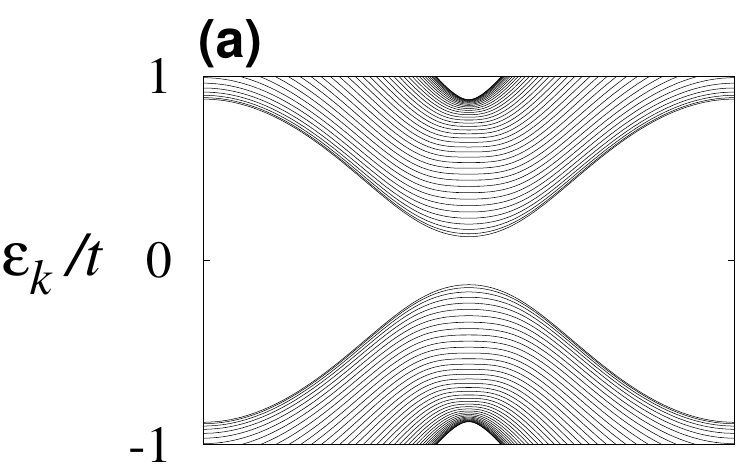}
		\includegraphics[width=5.0cm]{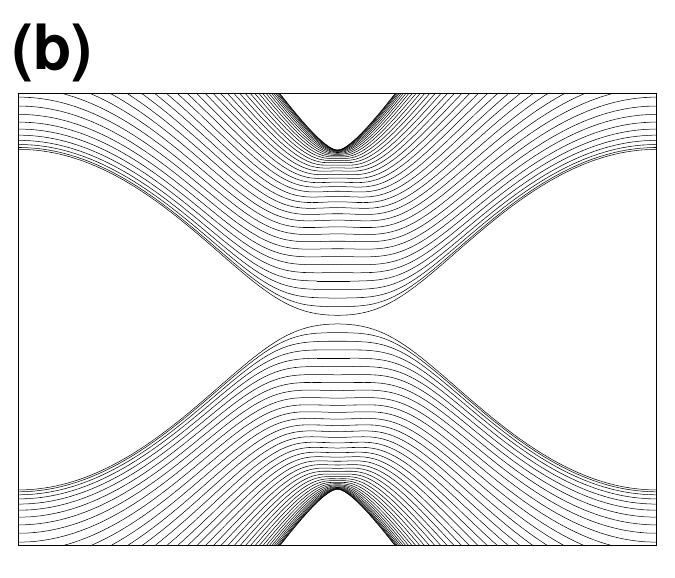}
		\includegraphics[width=5.0cm]{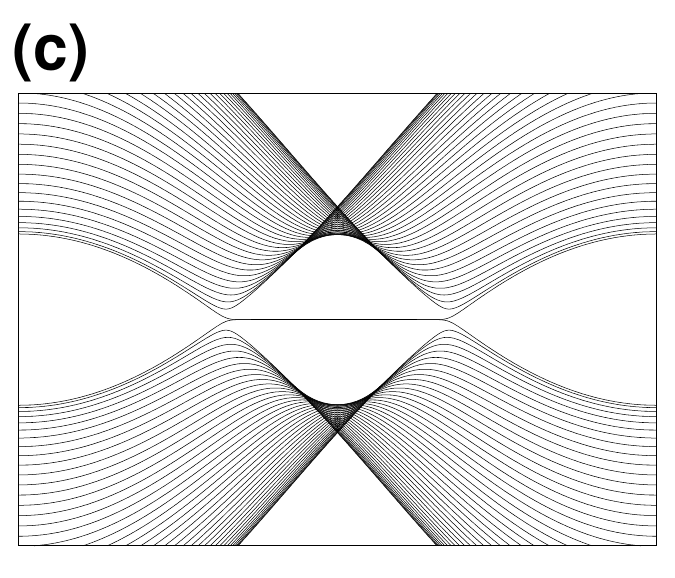}

		\includegraphics[width=6.86cm]{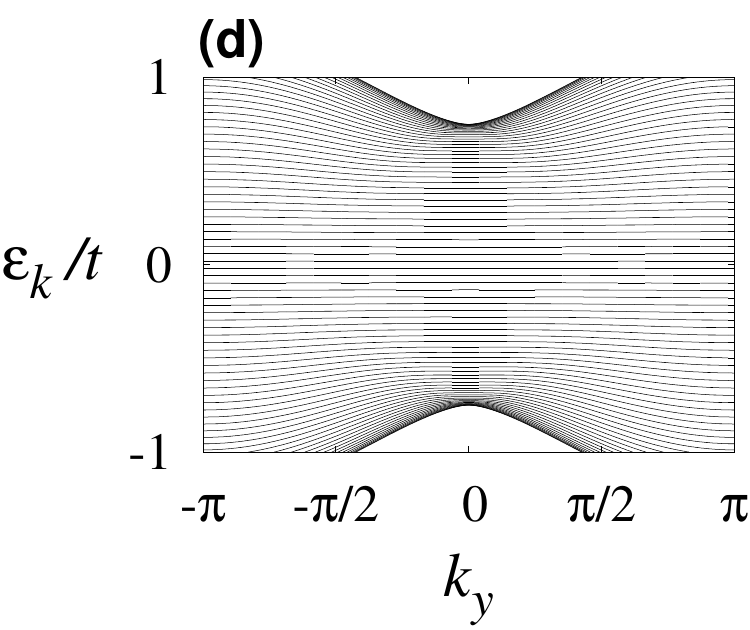}
		\includegraphics[width=5.0cm]{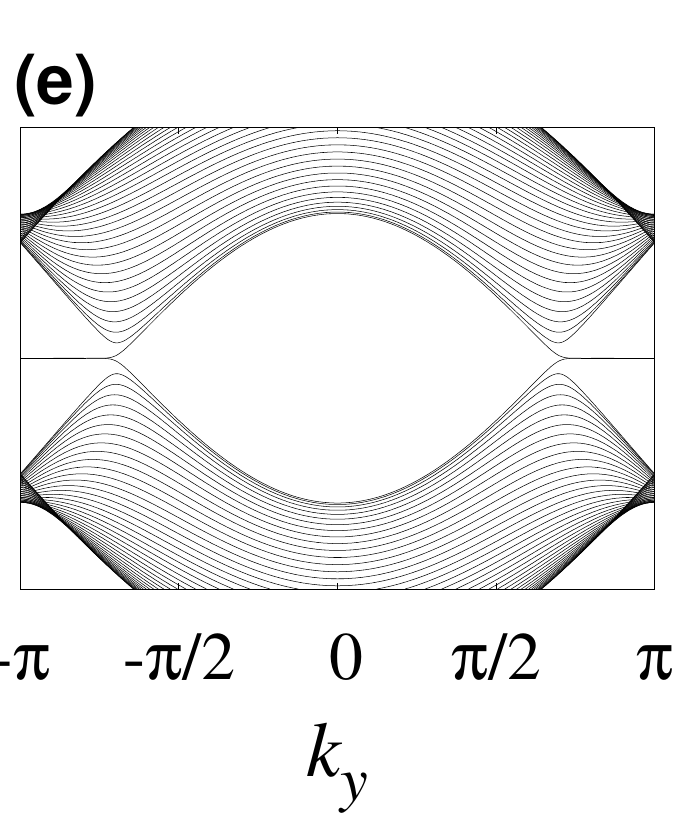}
		\includegraphics[width=5.0cm]{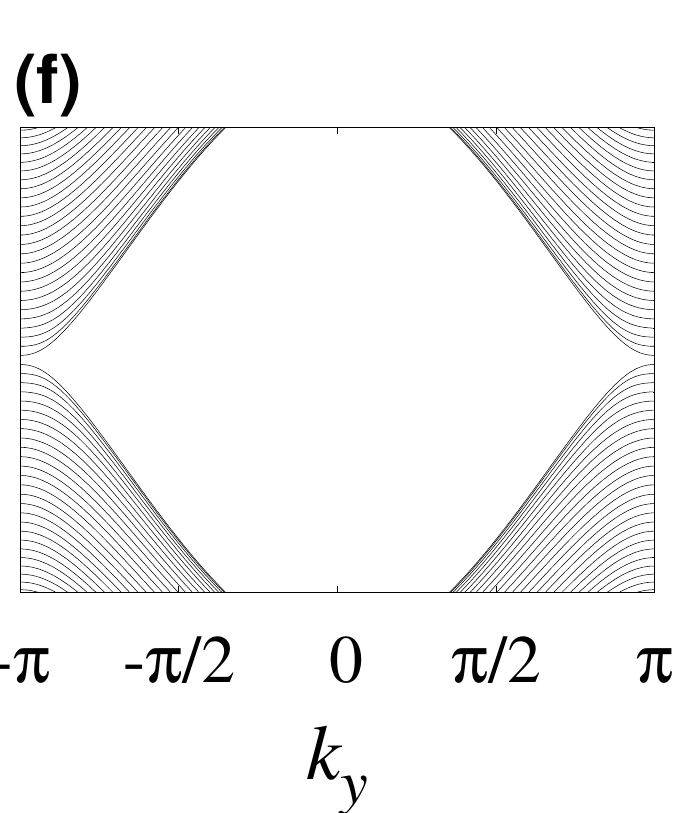}

		\caption{Energy dispersions of the modified square-lattice Kitaev model in
		  the ribbon of width $N_{x}=80$ for $\tdp=-0.25t$, $\Delta=t$ and 
		  $\mu=$ (a) $2t$; (b) $1.75t$; (c) $t$; (d) $0.25t$; (e) $-t$;
		  (f) $-2.25t$.
		  The energy dispersions displayed in this figure correspond 
		  to the empty circles at $\tdp=-0.25t$ in
		  Fig.~\ref{phase_diagram}.} \label{Kitaev_phase_5} 
\end{figure*}
we obtain the Hamiltonian of a block diagonal form

For this purpose, 
it is convenient to define two
Majorana operators given by
\begin{align}	
		  \nonumber 
		  a_{\bm{r}} &\equiv c^{\dagger}_{\bm{r}}+c_{\bm{r}},
	\\
		  b_{\bm{r}} & \equiv -i\left(c^{\dagger}_{\bm{r}}-c_{\bm{r}}\right),
	\label{defi_Mj}
\end{align}
which satisfy
the anticommutation relations 
\begin{align}
	\nonumber
		   \{a_{\bm{r}}, a_{\bm{r}'}\} &=2\delta_{\bm{r},\bm{r}'}, 
\\ 
	\nonumber
		   \{b_{\bm{r}}, b_{\bm{r}'}\} &=2\delta_{\bm{r},\bm{r}'},
\\
		   \{a_{\bm{r}}, b_{\bm{r}'}\} &=0. 
	\label{cr_Mj}
\end{align}
Inserting Eq. (\ref{defi_Mj}) into Eq. (\ref{H}), we obtain the Hamiltonian of the form
\begin{align}
	\nonumber
H=\frac{i}{4}\sum_{\bm{r}}&
\bigg[\left(t+\Delta\right)\sum_{\bm{a}}a_{\bm{r}}b_{\bm{r}+\bm{a}}
+\left(t-\Delta\right)\sum_{\bm{a}}a_{\bm{r}+\bm{a}}b_{\bm{r}}
\\
&+\,\tdp a_{\bm{r}}b_{\bm{r}+\bm{b}^{\prime}}+\,\tdp a_{\bm{r}+\bm{b}^{\prime}}b_{\bm{r}}-2\mu a_{\bm{r}}b_{\bm{r}}-h.c.\bigg].
	\label{eq_Mj_H}
\end{align}
We employ
open boundary conditions in the $x$ direction and periodic boundary conditions
in the $y$ direction. 
The $N_{x}\times N_{y}$ lattice sites in the ribbon can be denoted by $\bm{r}=m
a\hat{\bm{i}}+n a\hat{\bm{j}} ( m = 1,2,\cdots, N_x, n = 1,2, \cdots N_y)$. 
Then the Hamiltonian in the ribbon geometry can be written
\begin{align}
\nonumber
H_{N_{x}}=
\frac{i}{4}\sum^{N_{x}}_{m=1}\sum^{N_{y}}_{n=1}
		  &\big[\left(t-\Delta\right)\left(a_{m,n+1}b_{m,n}+a_{m+1,n}b_{m,n}\right)
\\
&+\left(t+\Delta\right)\left(a_{m,n}b_{m,n+1}+a_{m,n}b_{m+1,n}\right)
\nonumber
\\
\nonumber
		  &+\tdp ( a_{m,n}b_{m+1,n-1} + a_{m+1,n-1}b_{m,n} )
\\
&-2\mu a_{m,n}b_{m,n}-h.c.\big]
	\label{eq_H_Mj_OBC}
\end{align}
with the assumptions 
\begin{align}
		  \nonumber
		  a_{m,N_y{+}1} &= a_{m,1},
		  \\ 
		  \nonumber
		  b_{m,N_y{+}1} &= b_{m,1},
		  \\
		  a_{N_x{+}1,n} &= b_{N_x{+}1,n} =0. 
\end{align}
\begin{figure*}
		
		\includegraphics[width=6.56cm]{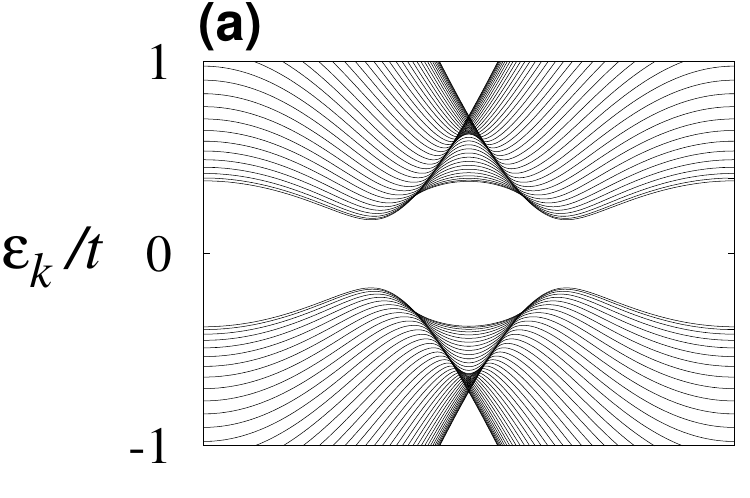}
		\includegraphics[width=5.0cm]{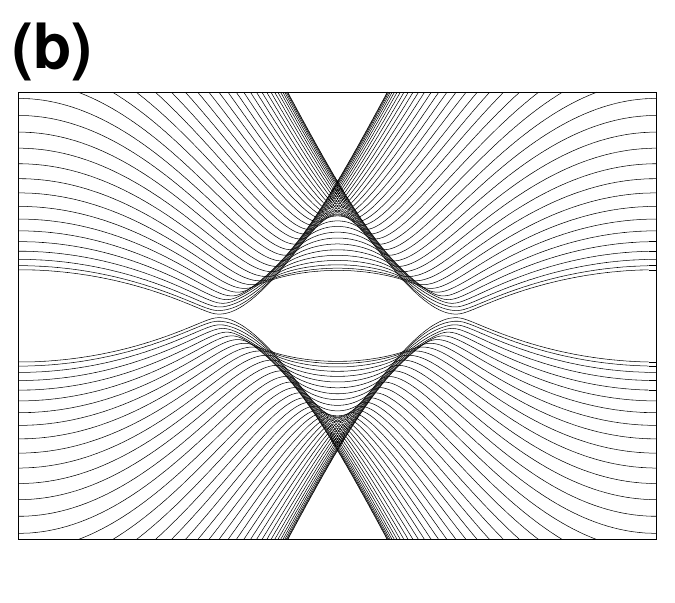}
		\includegraphics[width=5.0cm]{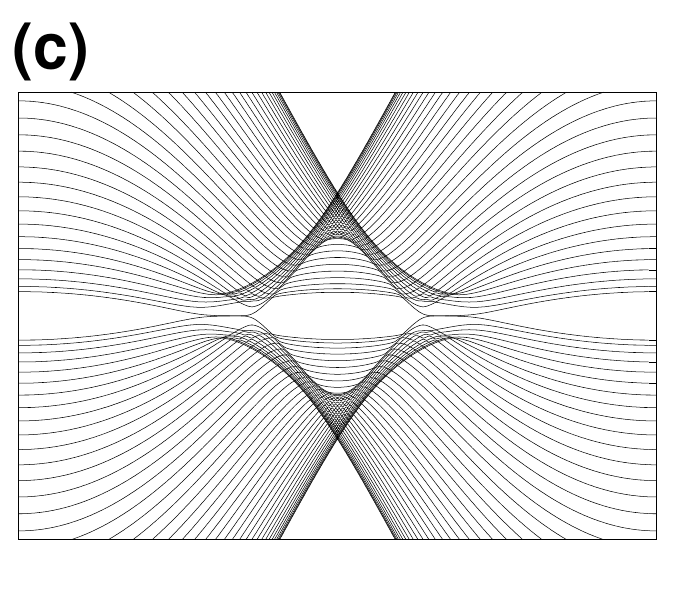}
		
		\includegraphics[width=6.56cm]{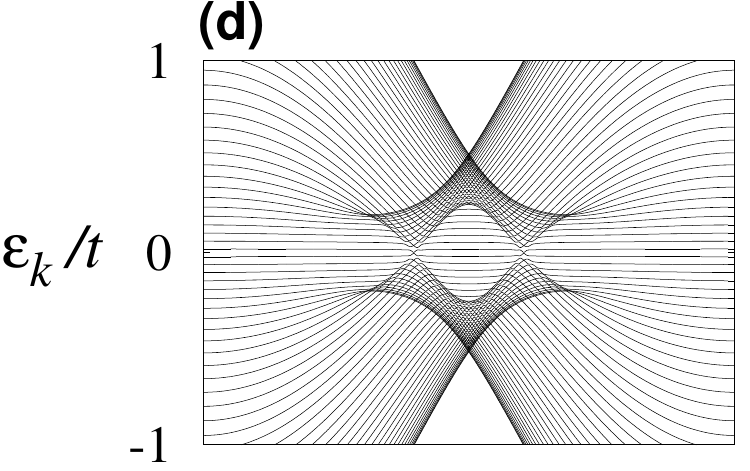}
		\includegraphics[width=5.0cm]{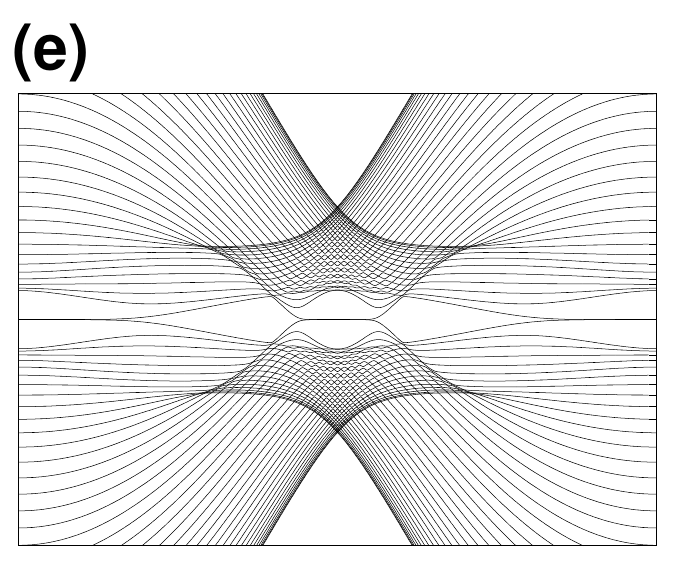}
		\includegraphics[width=5.0cm]{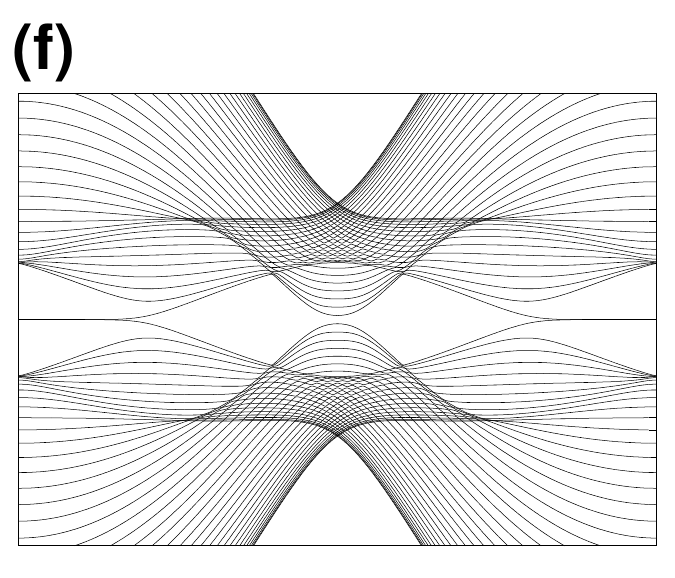}
		
		\includegraphics[width=6.56cm]{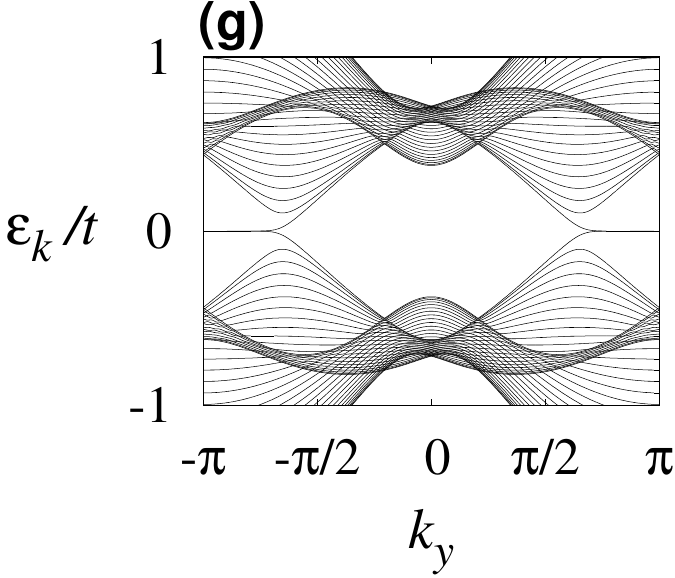}
		\includegraphics[width=5.0cm]{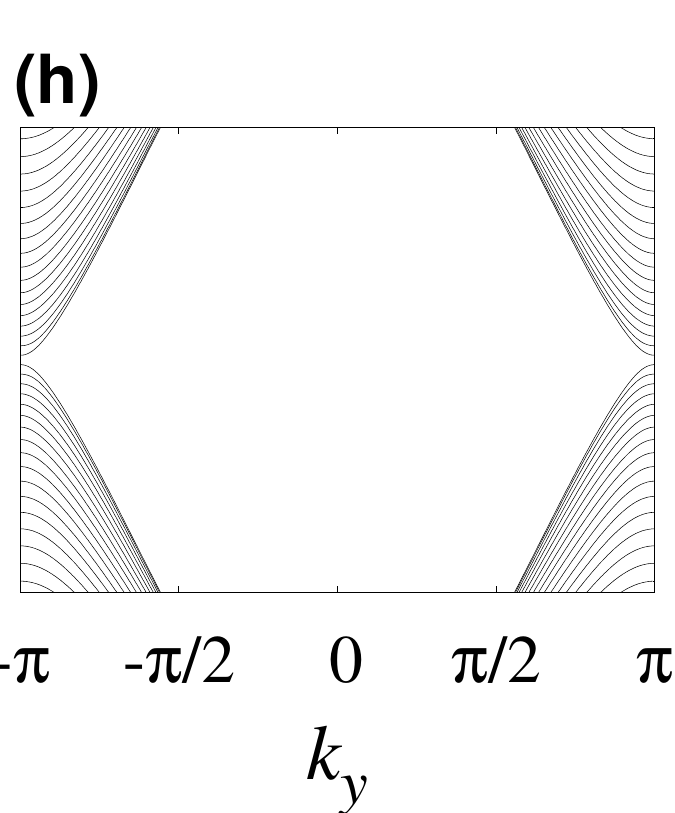}
		\includegraphics[width=5.0cm]{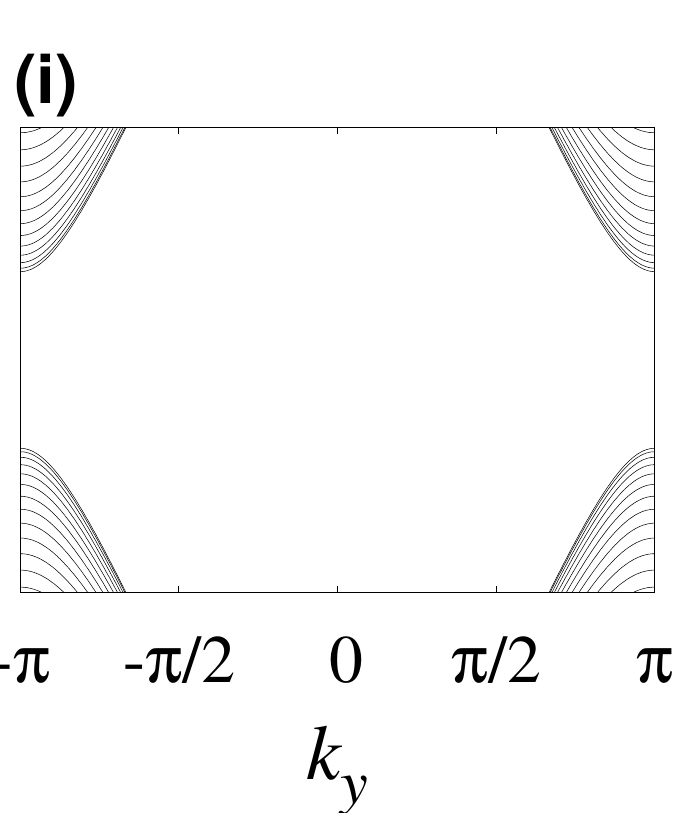}

		\caption{Energy dispersions of the modified square-lattice Kitaev model in
		  the ribbon of width $N_{x}=80$ for $\tdp=-0.25t$, $\Delta=t$, and
		  $\mu=$ (a) $2t$; (b) $1.65t$; (c) $1.45t$; (d) $1.25t$; (e) $t$;
		  (f) $0.75t$; (g) $0$; (h) $-3.25t$; (i) $-4t$. 
		  The energy dispersions displayed in this figure correspond to the
		  filled circles at $\tdp=-1.25t$ in Fig.~\ref{phase_diagram}.}
		  \label{Kitaev_phase_6} 
\end{figure*}

Taking Fourier transformation of Majorana operators in the $y$ direction
\begin{eqnarray}
	\nonumber
	a_{m,n}&=&\displaystyle\frac{1}{\sqrt{N_{y}}}\sum_{k_{y}}e^{ik_{y}n}\alpha_{m,k_{y}},
	\\ b_{m,n}&=&\displaystyle\frac{1}{\sqrt{N_{y}}}\sum_{k_{y}}e^{ik_{y}n}\beta_{m,k_{y}},	
	\label{ft_Mj}
\end{eqnarray}
%
\begin{align}
	\nonumber
	H_{N_{x}}=\sum_{k_{y}}&H^{k_{y}}_{N_{x}}
	\label{eq_df_H_Mj}
\end{align}
with
\begin{align}
	\nonumber
H^{k_{y}}_{N_{x}}=\sum^{N_{x}-1}_{m=1}\big[&\eta\,\alpha_{m,k_{y}} i \beta_{m,-k_{y}}+\delta_{1D}\,\alpha_{m+1,k_{y}}i\beta_{m,-k_{y}}
\\
\nonumber
&+\delta_{2D}\,\alpha_{m,k_{y}}i\beta_{m+1,-k_{y}}\big]
\\
&+\eta\,\alpha_{N_{x},k_{y}}i\beta_{N_{x},-k_{y}}+h.c.,
\end{align}
where
\begin{align}
	\nonumber
	&\eta\equiv\frac{1}{4}[\left(t-\Delta\right)e^{ik_{y}}+\left(t+\Delta\right)e^{-ik_{y}}-2\mu],
	\nonumber
	\\ &\delta_{1D}\equiv\frac{1}{4}\left(t-\Delta+\tdp e^{-ik_{y}}\right),
	\nonumber
	\\ &\delta_{2D}\equiv\frac{1}{4}\left(t+\Delta+\tdp e^{ik_{y}}\right).
\end{align}
We calculate the eigenvalues of $ H^{k_{y}}_{N_{x}} $ numerically to obtain the energy
spectra for $k_{y}$, which yields the energy dispersions in the ribbon.
We plot the energy dispersions of two typical cases, $\tdp=-0.25t$ and
$\tdp=-1.25t$, in Figs.~\ref{Kitaev_phase_5} and \ref{Kitaev_phase_6},
respectively.  

For $-\frac{1}{2}t<\tdp<0$, we find that Majorana zero modes show up when a
gapless point splits into two Weyl points. 
In Fig.~\ref{Kitaev_phase_5}(c), we observe Majorana zero modes in the interval,
the end points of which are the projections of 
two bulk Weyl points on the $k_{y}$ axis.
At $\mu=-\tdp$, where a nodal line exists in the bulk, the ribbon becomes
metallic.
For lower chemical potentials, the ribbon still exhibits a Majorana flat band,
whose interval contains $k_y=\pi$ instead of $k_y=0$.
As the chemical potential decreases further, the interval of Majorana band
shrinks gradually, and the flat band disappears 
when a pair of Weyl points merge into a point (Fig.~\ref{Kitaev_phase_5}(f)). 

For $\tdp<-\frac{1}{2}t$, the remarkable feature is that two Majorana flat bands 
exist in the energy dispersions (Fig.~\ref{Kitaev_phase_6}(c)); 
this has its origin in the fact that the system has four Weyl points which 
are generated from two separate trivial gapless points in the bulk.
The ribbon also exhibits metallic dispersions at $\mu=-\tdp$ (Fig.~\ref{Kitaev_phase_6}(d)), 
and the reduction of the chemical potential below this value changes the region of 
zero-energy edge modes. 
One interval contains $k_y=0$ and the other $k_y=\pi$ (Fig.~\ref{Kitaev_phase_6}(e)). 
The one with $k_y=0$, which connects the inner pair of Weyl points, first disappears 
(see Fig.~\ref{Kitaev_phase_6}(f)), and 
for $\mu=-2t+\tdp$ the other finally disappears (see Fig.~\ref{Kitaev_phase_6}(h)). 

The analytical approach for the existence of Majorana zero-energy edge modes also
sheds some insights on our understanding.
We express $ H^{k_{y}}_{N_x} $ in terms of a vector
$V_{k_{y}} \equiv \left(\alpha_{1,k_{y}}\;\,\beta_{1,k_{y}}\;\,\alpha_{2,k_{y}}\;\,\beta_{2,k_{y}}\;\cdots\;\alpha_{N_x,k_{y}}\;\,\beta_{N_x,k_{y}}\right) $, leading to
\begin{align}
	\begin{matrix}
		H^{k_{y}}_{N_x}=V_{k_{y}}\,h^{k_{y}}_{N_x}\,V^{\dagger}_{k_{y}}.
		\label{eq_H_Mj_matrix}
	\end{matrix}
\end{align}
When we impose that the eigenenergy be exactly zero, we can find that 
the corresponding eigenstates
of $ h^{k_{y}}_{N_x} $ which satisfy the open boundary conditions in the $x$
direction can exist only in two ways;
for $|p_{\pm}|<1$
\begin{eqnarray}
	\gamma_{k_{y}}^{(1)}&=&A_1\sum_{m=1}^{N_x}\left[(p_{+})^{m}-(p_{-})^{m}\right]\beta_{m,k_{y}},
	\label{eq_Kitaev_zeromodes1}
	\\
	\nonumber
	\gamma_{k_{y}}^{(2)}&=&A_2\sum_{m=1}^{N_x}\left[({p_{-}}^{*})^{N_x+1-m}-({p_{+}}^{*})^{N_x+1-m}\right]\alpha_{m,k_{y}}
	\\
\end{eqnarray}
and 
for $|p_{\pm}|>1$
\begin{eqnarray}
	\gamma_{k_{y}}^{(1)}&=&A_3\sum_{m=1}^{N_x}\left[({p_{-}}^{*})^{-m}-({p_{+}}^{*})^{-m}\right]\alpha_{m,k_{y}}.
	\\
	\nonumber
	\gamma_{k_{y}}^{(2)}&=&A_4\sum_{m=1}^{N_x}\left[(p_{+})^{m-N_x-1}-(p_{-})^{m-N_x-1}\right]\beta_{m,k_{y}},
	\\
	\label{eq_Kitaev_zeromodes2}
\end{eqnarray}
where
\begin{equation}
	p_{\pm} \equiv \frac{-\eta\pm\sqrt{\eta^{2}-4\delta_{1D}\delta_{2D}}}{2\delta_{2D}}
	\label{rate}
\end{equation}
and $A_i$'s are normalized constants.
Majorana zero-energy eigenstates $\gamma_{k_y}^{(1)}$  and $\gamma_{k_y}^{(2)}$
are localized around $m=1$ and $m=N_x$, respectively, and satisfy the boundary
condition at the opposite edge only in the limit of infinite width of the ribbon.

The necessary and sufficient condition for the existence of Majorana zero-energy modes 
is given by
\begin{eqnarray}
	\left(|p_{+}|-1\right)\left(|p_{-}|-1\right)>0.
	\label{neq_condition}
\end{eqnarray}
Equation~(\ref{neq_condition}) combined with Eq.~(\ref{rate}) 
leads to an inequality for $k_y$
\begin{eqnarray}
	\left(\mu+\tdp\right)\left(-2\tdp\cos^{2}{k_{y}}-2t\cos{k_{y}}+\mu+\tdp\right) < 0.
		  \label{MJband}
\end{eqnarray}
We verified that 
the intervals of Majorana flat bands that the inequality in Eq.~(\ref{MJband}) produces
are in complete agreement with the numerical results in 
Figs.~\ref{Kitaev_phase_5} and \ref{Kitaev_phase_6}. 
Particularly, the abrupt inversion of flat bands which occurs at $\mu=-\tdp$ can
be attributed to the prefactor $\mu + \tdp$ in Eq.~(\ref{MJband}).

It is also remarkable that
Eqs.~(\ref{eq_Kitaev_zeromodes1})-(\ref{eq_Kitaev_zeromodes2}) predict
which type of Majorana fermions exists at the edges.
According to Eqs.~(\ref{eq_Kitaev_zeromodes1})-(\ref{eq_Kitaev_zeromodes2}), 
in the region with $|p_\pm|<1$ Majorana fermions of a $\beta$ type exist at
the edges of $m=1$ and $\alpha$-type Majorana fermions around $m=N_x$. 
For $|p_\pm|>1$, on the other hand, $\alpha$-type ones at $m=1$ while
$\beta$-type ones at $m=N_x$.
We observe that 
these predictions agree well with numerical results presented above.
In the energy dispersion of Fig.~\ref{Kitaev_phase_6}(c), 
the intervals of both flat bands satisfy $|p_{\pm}|<1$. 
Numerically computed eigenstates demonstrate that these bands are degenerate with
two eigenstates; one with only $\alpha$-type components occupied is localized around 
$m=N_x$ and 
the other with only $\beta$-type components occupied is localized around $m=1$.
We can find different cases in Fig.~\ref{Kitaev_phase_6}(e), 
The inner flat bands containing $k_y=0$ satisfy $|p_{\pm}|<1$ and belong to the same
class of the flat bands described above. 
The flat band around $k_y=\pi$ with $|p_{\pm}|>1$, on the other hand, 
turns out to produce $\alpha$-type Majorana fermions around $m=1$ and
$\beta$-type ones around $m=N_x$.
and this reproduces well the prediction by the above analytical approach.

\section{Summary}

In this paper, we have considered the modified square-lattice Kitaev model, paying attention to
the effects of additional next-nearest-neighbor hopping on 
topological gapless phases and Majorana zero-energy edge modes. 
In addition to the topological gapless phase with two gapless points, which the original model exhibits 
in the presence of nearest neighbor hopping only, we have discovered newly emergent 
topological phases with four gapless points in the bulk.

The variation of the positions and the topological characters of the gapless points 
has been explored in detail
as chemical potential changes in the bulk energy spectra.
We have also
investigated the evolution of Majorana zero modes in the energy dispersions of
the ribbon geometry. 

Particularly, 
we have derived the analytical expression which determines the location of the Majorana flat
bands from the constraints on the boundary conditions of zero-energy edge modes;
this is in complete agreement with all the numerical results.
A detailed analytical analysis on the eigenstates corresponding to Majorana
zero-energy modes has provided deeper understanding on 
the presence of two types of Majorana fermion 
which has its origin in topological features of the system.

\begin{acknowledgements}
	This work was supported by the National Research Foundation of Korea through Grant No. NRF-2021R1F1A1062773.
\end{acknowledgements}

\bibliography{Kitaev}

\end{document}